\documentclass[12pt]{article}%
\usepackage{amsfonts}
\usepackage{amssymb}
\usepackage{graphicx}
\usepackage{setspace}
\usepackage[round]{natbib}
\usepackage{amsmath}%
\usepackage{amsthm}%
\usepackage{palatino}
\usepackage{paralist}
\usepackage{multirow}
\usepackage{booktabs}
\usepackage{dsfont}
\usepackage{indentfirst}
\usepackage{enumerate}
\usepackage{longtable}
\usepackage{caption}
\usepackage{subcaption}
\usepackage{mwe}
\usepackage[utf8]{inputenc}
\usepackage[T1]{fontenc}
\usepackage[hyperindex,breaklinks]{hyperref}
\hypersetup{colorlinks=true,       % false: boxed links; true: colored links
	linkcolor=red,       
	citecolor=blue,        % color of links to bibliography
	filecolor=magenta,      % color of file links
	urlcolor=cyan           % color of external links
}  

\usepackage{appendix}                      
\usepackage{lscape}       
\usepackage{comment}       
\newtheorem{proposition}{Proposition}      
\usepackage{algorithm}                     
\usepackage{algpseudocode}                 
\DeclareMathOperator*{\argmax}{arg\,max}  
   
\usepackage{tikz}
\usetikzlibrary{patterns} % For dash patterns
\usepackage{changepage} 
\usepackage{setspace}
\vspace{-5cm}
\title{Stochastic Discount Factors with Cross-Asset Spillovers
     \thanks{
		     \footnotesize
		        We are grateful to Utpal Bhattacharya, Lin Will Cong, Yi Ding, Gavin Feng, Shuyi Ge, Michael Gofman, Jingyu He, Kewei Hou, Wenjin Kang, Shikun Barry Ke, Junye Li, Sicong Li, Xin Liu, Semyon Malamud, Stefan Nagel, Zilong Niu, Alex Philipov, Gil Segal, Giorgia Simion, Robert Stambaugh, Yinan Su, Dragon Yongjun Tang, Jun Tu, Siwei Wang, Yanchu Wang, Dacheng Xiu, Jingzhou Yan, Jingyi Yao, Jun Yu, Chao Zhang, Dake Zhang, Guofu Zhou, Qi Zhou,
		     and seminar and conference participants at 
             Hebrew University of Jerusalem,
             Shanghai Jiao Tong University, 
             Sichuan University, Southwestern University of Finance and Economics, University of Science and Technology of China, Xiamen University,
             Fudan International Symposium on AI in Finance 2025,
             Hong Kong Conference for Fintech, AI, and Big Data in Business 2025,
             Paris December Finance Meeting 2025,
             SYSU Conference on Big Data, AI, and FinTech 2025,
             UMacau FinTech and Financial Markets Workshop 2025,
             and
             USTC Frontiers in Finance Conference 2025,
             for constructive discussions and feedback. 
		     The authors acknowledge financial support from 
		     Inquire Europe.
             
			Avramov (E-mail: \texttt{doron.avramov@runi.ac.il}) is at  Reichman University (IDC), Herzliya, Israel.
			He (E-mail: \texttt{xin.he@ustc.edu.cn}) is at University of Science and Technology of China.
            All authors contributed equally to this work.

		}
        
}

\author{
    Doron Avramov \\ 
    \textit{\small Reichman University (IDC), Herzliya, Israel}
	\and 
	Xin He \\ 
    \textit{\small University of Science and Technology of China}
}

\date{\today}

\begin{document}

    \clearpage
    
    \maketitle
    
	\thispagestyle{empty}
	
    \vspace{-1cm}
    
	\begin{abstract}

This paper develops a unified framework that links firm-level predictive signals, cross-asset spillovers, and the stochastic discount factor (SDF). Signals and spillovers are jointly estimated by maximizing the Sharpe ratio, yielding an interpretable SDF that both ranks characteristic relevance and uncovers the direction of predictive influence across assets. Out-of-sample, the SDF consistently outperforms self-predictive and expected-return benchmarks across investment universes and market states. The inferred information network highlights large, low-turnover firms as net transmitters. The framework offers a clear, economically grounded view of the informational architecture underlying cross-sectional return dynamics.

		\bigskip
		
		
		\bigskip
		
		
	\end{abstract}
	
	% \newpage

 %    \thispagestyle{empty}

 %    \begin{center}
 %        \textbf{Conflict-of-interest disclosure statement}    
 %    \end{center}
    
 %    Doron Avramov
    
 %    I have nothing to disclose.
    
 %    Xin He
    
 %    I have nothing to disclose.
    
    \newpage
	
	\setcounter{page}{1}

 \section{Introduction} 
 \label{sec:introduction}
The central objective of empirical asset pricing is to identify firm-level signals that explain the cross-section of expected stock returns—whether through exposure to risk factors or persistent mispricing. The dominant paradigm, grounded in the assumption of self-predictability, asserts that a firm’s own characteristics forecast its own returns (see, e.g., \cite{cochrane2011presidential, harvey2016and}). Complementing this view is a growing literature on cross-predictability—the idea that the characteristics or returns of one asset can help forecast the returns of others (see, e.g., \cite{lo1990contrarian, hou2007industry, cohen2008economic, cohen2012complicated, huang2021psychological, huang2022frog}). A key mechanism underpinning this phenomenon is the presence of lead–lag effects, whereby price movements or information from one firm precede and predict those of related firms. Such effects can stem from staggered information diffusion, peer influence within industries, supply chain linkages, or correlated trading by institutional investors that induces price pressure across related assets.

Despite recent methodological advances in modeling cross-stock predictability, several foundational questions remain unresolved. Chief among them is how a mean–variance investor can analytically integrate multiple predictive signals when returns are interconnected across assets. Equally crucial is developing a framework that jointly captures both the relevance of individual signals and the structure of return spillovers—enhancing portfolio performance while preserving interpretability.

This paper addresses these questions by proposing a unified and systematic framework for constructing maximum–Sharpe ratio strategies. We combine firm-level signals through a flexible weighting vector (the signal-aggregation vector $\Lambda$) and model cross-asset spillovers using a structured connection matrix (the spillover matrix $\Psi$). The resulting optimal strategy admits a transparent analytical characterization. This formulation naturally connects to the stochastic discount factor (SDF; see \cite{hansen1991implications, cochrane2009asset, back2017asset}), which, in this context, takes the form of a single factor that prices the cross-section of returns.

An important distinction in the asset pricing literature lies between conditional and unconditional Sharpe ratio optimization. As emphasized by \cite{hansen1987role}, conditional optimization targets the best return–risk trade-off at each point in time using the information then available, whereas unconditional optimization maximizes this trade-off in expectation using long-run moments.\footnote{See also \citet{lewellen2006}, who emphasize the distinction between conditional and unconditional beta pricing.}
Our framework follows the latter approach: while it incorporates time-varying signals—such as firm characteristics and cross-asset linkages—the stochastic discount factor is optimized to perform well on average over time. This orientation prioritizes long-horizon performance over period-by-period efficiency, yielding strategies that are transparent, robust, and empirically grounded.

While the analytical formulation provides a population-level characterization of the Sharpe-optimal SDF, our empirical implementation uses a regression-based procedure tailored for high-dimensional applications. We build on the approach of \cite{britten1999sampling} and employ ridge-type regularization—with a single tuning parameter $\lambda$ chosen by five-fold cross-validation—to estimate both the signal weights and the connection matrix. This method converges to the theoretical solution in large samples while enhancing numerical stability and interpretability. Unlike expected return-maximization—which, under certain specifications, can lead to extreme concentration in a single predictor—Sharpe ratio-maximization encourages diversification across signals, thereby enhancing robustness and practical relevance.

To build intuition, we start with a low-dimensional toy example using five well-known firm characteristics and nine portfolios sorted by size and book-to-market. This simplified setting enables us to illustrate the estimated signal weights, cross-asset linkages, and resulting trading strategy in full detail. We evaluate performance with a rolling out-of-sample procedure, re-estimating the strategy each month using the prior 10 years of data. Even in this controlled environment, the maximum–Sharpe ratio strategy based on cross-stock predictability attains an annualized Sharpe ratio of 1.22, compared with 0.60 for the self-predictive benchmark—an improvement driven simultaneously by cross-asset spillovers, shifts in signal relevance, and their interaction.

We then scale the framework to a comprehensive empirical setting using 138 firm-level signals from the \cite{jensen2023there} dataset. Our primary investment universe consists of 138 univariate spread portfolios spanning 1963–2023. We also consider a broader set of 544 bivariate portfolios sorted by firm size and a secondary characteristic. Applying the same rolling 10-year estimation scheme, the maximum–Sharpe ratio (MS) strategy attains annualized Sharpe ratios of 2.21 and 3.32 on the spread and bi-sort portfolios, respectively—consistently outperforming both self-predictive benchmarks and maximum-expected return (MR) strategies.
Specifically, the Sharpe ratio of our cross-predictive SDF strategy exceeds that of a self-predictive Sharpe ratio–maximizing benchmark by 0.79 on spread portfolios and more than 1.26 on bi-sorted portfolios—translating into economically meaningful gains in certainty-equivalent returns. Moreover, compared to expected return–maximizing strategies, our Sharpe ratio–maximizing SDF improves risk-adjusted performance by factors of 4–10, depending on the investment universe and market regime.

To assess robustness, we evaluate performance across different market environments. We split the test sample by investor sentiment and by volatility regimes based on the VIX index. The Sharpe ratio–maximizing strategy maintains strong performance across all subsamples. For example, in high-sentiment periods, the strategy delivers a Sharpe ratio of 2.19 on spread portfolios and 3.58 on bi-sort portfolios. Even in low-sentiment or high-volatility regimes—conditions that typically challenge individual anomaly-based strategies—the strategy sustains Sharpe ratios above 2. These results contrast with the more state-dependent performance of expected return–maximizing portfolios.

The SDF defines a single factor that, ex ante, prices the cross-sectional variation in expected returns of the test assets. We evaluate whether this factor's payoffs are priced by leading asset pricing models and find sizable, statistically significant alphas relative to a broad set of benchmarks. These include the liquidity factor \cite{pastor2003}, the Fama–French five-factor model \cite{fama2015five}, the q-factors \cite{hou2015}, the mispricing factors \cite{stambaugh2017}, the behavioral factors \cite{daniel2020short}, and a comprehensive fourteen-factor model. Across all specifications, the strategy delivers alphas of about 0.25\% per month with $t$-statistics above 11, indicating that the return variation embedded in cross-asset spillovers is not captured by existing models.

Upon optimizing the Sharpe ratio, we uncover the underlying economic drivers of return predictability. By examining the estimated weights assigned to firm-level characteristics, we find that the most influential predictors cluster in the categories of investment, value, and profitability, with signals such as liquidity of book assets, dividend yield, and return on equity consistently receiving the highest weights. In contrast, return-based signals—including momentum, short-term reversal, and seasonality—exhibit persistently low weights. This pattern suggests that the cross-predictive SDF is primarily anchored in stable firm fundamentals rather than transitory market signals.

In optimizing the Sharpe ratio, we also obtain a connection matrix, denoted by $\Psi$, that encodes the predictive relationships across stocks. Each entry $\Psi_{i,j}$ reflects the extent to which signals from asset $i$ forecast the returns of asset $j$, while diagonal elements represent self-predictive strength. Empirically, the average off-diagonal entry is substantial—often exceeding the average diagonal—indicating that cross-asset predictive linkages carry more information than self-predictive signals alone. Aggregating rows and columns of the matrix following \cite{diebold2014network}, we uncover a directional structure: certain stocks consistently act as net transmitters of predictive signals, while others serve primarily as net receivers. Transmitters are typically large and low-turnover, whereas receivers tend to be smaller, high-turnover stocks with characteristics such as value orientation, high profitability, low investment activity, and strong past returns.

It is worth noting that the Sharpe ratio of the cross-predictive strategy is time-varying and declines notably after 2000. In the 1990s, the strategy delivers exceptional performance, with Sharpe ratios exceeding 2 on spread portfolios and above 4 on bi-sort portfolios. However, performance attenuates in the post-2000 period, mirroring the broader decline in self-predictability. For instance, \cite{green2017characteristics} document that many anomaly portfolios become less profitable after 2003, attributing the decline to the widespread adoption of anomaly-based strategies, improved market liquidity, and the growth of passive ETF investing.

Despite this attenuation, the proposed strategy maintains strong performance from 2000 to 2023, achieving Sharpe ratios of 1.58 (spread portfolios) and 2.21 (bi-sort portfolios)—substantially higher than those of standard benchmark factors: 0.41 (market), 0.27 (size), 0.20 (value), 0.54 (profitability), 0.43 (investment), and 0.09 (momentum). By the end of 2023, five-year trailing Sharpe ratios decline to approximately 1.2 for the spread and bi-sort strategies, yet both remain consistently superior to traditional benchmarks even in recent years.

The paper proceeds as follows. Section~\ref{sec:model} presents the econometric framework. Section~\ref{sec:estimation} outlines the estimation methodology. Section~\ref{sec:data} describes the data. Section~\ref{sec:results} reports the empirical findings. Section~\ref{sec:conclusion} concludes.

\section{Econometric Framework}
\label{sec:model}

We consider an investment universe consisting of \( N \) risky assets. At each time \( t \), the investor observes a signal matrix \( S_t \in \mathbb{R}^{N \times M} \), where each row corresponds to one asset and contains \( M \) predictive characteristics (e.g., size, valuation, profitability, investment, past returns). Each column of \( S_t \) is cross-sectionally standardized to have zero mean and unit variance. Although our framework allows for a time-varying number of assets, the empirical analysis focuses on a fixed cross-section of sorted portfolios. We define \( t = 1 \) as the first period in which signals are observed, and \( t = T + 1 \) as the final period in which asset returns are realized.

\subsection{Trading Strategy} \label{sec:linear_strategy}
A linear strategy that incorporates multiple signals and cross-predictability is specified as  
\begin{equation}
	\omega_t' = \Lambda'\,S_t'\,\Psi,
	\label{eqn:omega}
\end{equation}  
where \( \omega_t \in \mathbb{R}^N \) denotes the portfolio weights, \( \Lambda \in \mathbb{R}^M \) assigns loadings to each signal, and \( \Psi \in \mathbb{R}^{N \times N} \) encodes how signals from one asset influence positions across all assets. Specifically, the weight on asset \( i \) is determined by multiplying \( \Lambda' \), \( S_t' \), and the \( i \)th column of \( \Psi \), allowing all signals in \( S_t \) to contribute to each asset’s position. The element \( \Psi_{i,j} \) quantifies the predictive impact of asset \( i \)’s signals on asset \( j \).

Relative to \citet{brandt2009parametric}, who model portfolio weights as a function of firm-specific attributes, our framework generalizes the approach by allowing economically meaningful cross-asset spillovers to shape portfolio allocations. Moreover, although we focus on linear strategies, the framework readily accommodates nonlinear extensions by enriching the signal matrix with polynomial or Fourier-based transformations. 
For instance, one can construct an expanded signal matrix of dimension \( N \times MP \), where the first \( N \times M \) block corresponds to the original \( S_t \), the second to its elementwise square, and subsequent blocks to higher-order transformations up to the \( P \)th power. Importantly, such extensions preserve the dimension of the $\Psi$ matrix, while the 
$\Lambda$ vector expands accordingly to accommodate the enlarged set of predictors—including higher-order powers of the original signals. We leave the formal development and empirical implementation of such nonlinear extensions to future research.

We construct  managed-portfolio returns in excess of the risk-free rate by interacting future returns with the current values of predictive signals:
\begin{equation}
	\Pi_s = \bigl(I_N \otimes r_s\bigr)\,S_t, 
\end{equation}
where $\Pi_s$ is an $N^2 \times M$ matrix of managed-portfolio returns, $I_N$ is the $N \times N$ identity matrix, $r_s$ is a vector of $N$ excess returns realized at time $s > t$, and $\otimes$ denotes the Kronecker product.

The expected returns on these managed-portfolios are then defined as  $	\Pi = E\bigl[\Pi_s\bigr]$. Additionally, define
\begin{equation}
	\Phi = \mathrm{vec}\bigl(\Psi'\bigr),
	\label{eqn:Phi}
\end{equation}
so that \(\Phi\in\mathbb{R}^{N^2}\). The vectorized \(\Phi\) and the matrix $\Pi$ streamline later expressions for portfolio outcomes.

To aid interpretation, limit extreme equity positions, and stabilize estimation, we impose Euclidean norm constraints on key parameters. Specifically, we set
\begin{equation} \label{eqn:norm_constraint}
	\Lambda'\Lambda = 1,
	\quad
	\Phi'\Phi = 1,
\end{equation}
where the Euclidean norm constraint on the vector $\Phi$ is equivalent to a Frobenius norm constraint on the matrix $\Psi$. From a Bayesian perspective, these constraints correspond to zero-mean Gaussian priors on $\Lambda$ and $\Phi$, inducing ridge-type regularization that penalizes large parameter values.

Proposition \ref{prp:linear_metrics} formulates the realized return of the strategy in a convenient form, along with the expected return and Sharpe ratio. Appendix \ref{app:linear_metrics} provides the proof.

	\begin{proposition} \label{prp:linear_metrics}
		The investment metrics are as follows:
		\begin{itemize}
			\item The realized and expected returns can be expressed as
			\begin{align}
				&\pi_s = \Lambda'\Pi_s\Phi, \\
				\label{eqn:expected_return}
				&E(\pi_s) = \Lambda'\Pi\Phi.
			\end{align}

\item The square of the Sharpe Ratio (\(SR^2\)) is given by the following two equivalent expressions:
\begin{eqnarray}
	SR^2 &=& \frac{\Lambda' A_\Phi \Lambda}{\Lambda' B_\Phi \Lambda},  \label{eqn:SR} \\
	SR^2 &=& \frac{\Phi^{\prime} A_{\Lambda} \Phi}{\Phi^{\prime} B_{\Lambda} \Phi}. \label{eqn:SR_alternative}
\end{eqnarray}
Here, \(A_\Phi = \Pi' \Phi \Phi' \Pi\), \(B_\Phi = (\Phi' \otimes I_M) \Sigma_{\Phi} (\Phi \otimes I_M)\), \(\Sigma_{\Phi}\) is the covariance matrix of \(\text{vec}(\Pi_s')\), and \(I_M\) is the identity matrix of order \(M\). Similarly, \(A_{\Lambda} = \Pi \Lambda \Lambda^\prime \Pi^\prime\), \(B_{\Lambda}=\left(\Lambda^{\prime} \otimes I_{N^2}\right) \Sigma_{\Lambda} \left(\Lambda \otimes I_{N^2}\right)\), \(\Sigma_\Lambda\) is the covariance matrix of \(\text{vec}(\Pi_s)\), and \(I_{N^2}\) is the identity matrix of order \(N^2\). 

\end{itemize}
	
	\end{proposition}

We offer several remarks regarding Proposition~\ref{prp:linear_metrics}.

First, our empirical analysis primarily focuses on maximizing the Sharpe ratio, using expected return-maximization as a benchmark for comparison. While both objectives rely on the same expressions for expected returns, they lead to different optimal estimates for the signal weight vector $\Lambda$ and the vectorized connection matrix $\Psi$. In particular, expected return-maximization reduces to a bilinear optimization problem with closed-form solutions, whereas Sharpe ratio-maximization entails solving a generalized eigenvalue problem via an iterative procedure.

Importantly, maximizing the squared Sharpe ratio necessitates the use of both representations of the Sharpe ratio provided in Proposition~\ref{prp:linear_metrics} when estimating the optimal values of $\Lambda$ and $\Phi$. Explicit solutions for both the expected return and Sharpe ratio-maximization problems are presented later in the paper.

Second, Proposition~\ref{prp:linear_metrics} makes extensive use of the vectorized form of $\Psi$, which fully retains the cross-predictive structure embedded in $\Psi$. As a result, the information content relevant for cross-predictability is entirely preserved in $\Phi$, ensuring that the resulting strategy remains grounded in the same underlying predictive relationships.
    
Third, the expression for investment return offers an intuitive economic interpretation of our trading strategy. Recall that $\Pi$ denotes the matrix of managed-portfolio expected returns, with each of its $N^2$ rows representing the expected value of one asset’s return multiplied by one of the $M$ signals across the $N$ assets. Under the normalization $\mathbb{E}[S_t]=0$, $\Pi$ simplifies to the covariance matrix between future asset returns and contemporaneous signal values. If characteristic $m$ of stock $j$ helps predict the future return of stock $i$, the corresponding element of $\Pi$ will be nonzero, reflecting this predictability.

Thus, in this framework, $\Lambda$ assigns relative weights to signals, $\Phi$ encodes cross-asset interactions, and together they operate on the matrix $\Pi$ to optimize investment metrics.

 Fourth, the expected return of the trading strategy can alternatively be expressed as  
 \begin{equation}
 	E(\pi_s) = \sum_{m=1}^{M} \Lambda_m \mu_m,
 \end{equation}
 where \(\mu_m = \sum_{p=1}^{N^2} \Pi_{pm} \Phi_p\) represents a weighted combination of portfolio expected returns, with \(\Pi_{pm}\) denoting the expected return of the corresponding managed-portfolio and \(\Phi_p\) capturing the strength of the \(p\)-th relationship within the strategy. 

This expected-return expression is informative because it demonstrates that, whether subject to an $L_1$ constraint or left unconstrained, the optimal solution is a corner solution: the trading strategy is entirely driven by the predictor with the largest absolute value of~$\mu_m$, denoted predictor~$j$, with $|\Lambda_j| = 1$ and all other elements of~$\Lambda$ equal to zero. In contrast, under an $L_2$ constraint, the optimal $\Lambda$ (given $\Phi$) is proportional to the $M$-vector that collects the $\mu_m$ values. By comparison, Sharpe ratio-maximization effectively harnesses the benefits of diversification across predictors, assigning meaningful weight to multiple signals.

In the context of expected return-maximization, \citet{he2024PPMulti} extend the principal portfolios framework of \citet{kelly2023principal} from a single-signal to a multi-signal setting by introducing a three-dimensional prediction tensor. Our study should not be viewed as a multi-predictor extension of principal portfolios. Rather, we propose a framework that differs in both econometric structure and economic objective. From a modeling standpoint, we focus on a two-dimensional matrix $\Pi$, where one dimension captures multiple signals and the other encodes cross-predictive relationships across assets. From an economic perspective, the proposed methodology is explicitly designed to flexibly optimize the Sharpe ratio.

Fifth, the realized return $\pi_s$ of the maximum-Sharpe ratio portfolio is proportional to the stochastic discount factor (SDF), as implied by the fundamental asset pricing identity \citep{hansen1991implications, cochrane2009asset, back2017asset}:
\begin{equation}
	M_s = 1 - \omega' r_s, \quad \text{with} \quad \mathbb{E}[M_s r_s] = 0,
	\label{eqn:sdf}
\end{equation}
where $M_s$ denotes the pricing kernel and $\omega$ is the vector of slope coefficients.
Identifying the true $\omega$ is challenging in finite samples due to the ``limits to learning'' highlighted by \citet{didisheim2024apt}. While the literature has proposed various estimators of the SDF, our approach introduces a novel proxy that explicitly captures cross-asset spillovers, distinguishing it from prior work.

Sixth, \citet{kelly2023principal} focus on expected return-maximization and propose an alpha-beta decomposition: the antisymmetric and symmetric components of the prediction matrix yield the principal alpha and principal exposure portfolios, respectively. Although our expected return-maximizing strategy can be cast within this framework, our Sharpe ratio-maximizing strategy—by construction—excludes alpha, consistent with the SDF interpretation in Equation~\eqref{eqn:sdf}.

Empirically, we demonstrate that expected return-maximizing and Sharpe ratio-maximizing strategies—both accounting for cross-asset spillovers—lead to substantially different outcomes. The Sharpe ratio-maximizing strategy consistently delivers significantly higher Sharpe ratios across the full sample, as well as during both expansion and contraction periods.

Next, sorting assets by the estimated weights surfaces the portfolio’s informational backbone: it ranks assets by how much they raise the strategy’s risk-adjusted payoff. High-ranked (large-weight) assets are those that sharpen the payoff of the pricing kernel in three complementary ways: they carry economically meaningful fundamentals; they occupy advantageous positions in the web of cross-asset co-movements that let the portfolio harness spillovers; and they help balance the residual risks created elsewhere in the strategy. An asset can rank highly even if its own return is not strongly predictable—when it acts as a conduit that improves how the portfolio captures cross-asset structure or when it completes the diversification needed to express valuable payoff directions more cleanly. Lower-ranked assets contribute less to efficiency either because their information is largely redundant or because they add volatility without commensurate benefit.

Finally, as shown in Appendix~\ref{sec:dgp}, the connection matrix $\Psi$ closely aligns with the projection of stock returns onto the distinct elements of the signals.

    \subsection{Zero-Cost and Leverage Constraints} 
    \label{sec:zero_cost_and_leverage}

    % % % % % % % % % % % % % % % % % % 
    % % Zero-Cost and Propostion 2. % %
    % % % % % % % % % % % % % % % % % % 

   Up to this point, we have only imposed norm constraints on the strategy's positions. However, empirical asset pricing typically requires a trading strategy, factor, or anomaly to take the form of a long-short portfolio—that is, to be zero-cost with total leverage equal to two.
   
   The following proposition imposes this zero-cost constraint on the strategy.

	\begin{proposition}
		A zero-cost trading strategy can be expressed as follows:
		\begin{eqnarray}
			\omega_t' &=& \Lambda' S_t' \Psi - \frac{1}{N} \Lambda' S_t' \Psi A, \\
			&=& \Lambda' S_t' \Psi \Theta,
		\end{eqnarray}
		where \(A\) is an \(N \times N\) matrix, with each element set to one, and \(\Theta\) = \(I_N - \frac{1}{N}A\).
	\end{proposition}

    Notice that \( \omega_t' \iota_N = 0 \), where \( \iota_N \) is an \(N\)-vector of ones. Fortunately, all previous derivations remain valid under the zero-cost constraint.
    
    The necessary modifications are as follows. Define \( \Pi_{si} = \Theta (r_{s} S_{it}') \) for each \( i = 1, 2, \ldots, N \), and construct \( \Pi_s \) by vertically stacking \( \Pi_{si} \). All investment metrics in Proposition \ref{prp:linear_metrics} can then be re-derived under the zero-cost constraint.
    
    In Appendix \ref{app:zero_cost}, we demonstrate that the zero-cost constraint reduces the expected profitability of the trading strategy. However, this constraint is essential for ensuring comparability across strategies.
    
   In our empirical analyses, we primarily focus on zero-cost strategies, where the long and short positions are of equal magnitude by construction. To further ensure comparability, we rescale these positions so that total portfolio leverage equals two. This adjustment aligns our strategies with standard practice in the literature \citep[e.g.,][]{fama1993common}.

 \section{Estimating the Unknown Parameters} 
 \label{sec:estimation}
 
We provide methods for estimating the unknown parameters underlying the trading strategy. 
 
 \subsection{Maximizing Expected Return}
 
The following proposition presents the solution for the strategy that maximizes expected return. 

 \begin{proposition}
 	\label{prp:max_expected_return}
 	By the Singular Value Decomposition (SVD), $\Pi$ can be decomposed as
 	\begin{equation}
 		\Pi = U \Lambda_{\Pi} V',    
 	\end{equation}
 	where \(U\) is an \(N^2 \times N^2\) orthogonal matrix, \(\Lambda_{\Pi}\) is an \(N^2 \times M\) diagonal matrix of singular values, and \(V'\) is an \(M \times M\) orthogonal matrix. 
 	
The estimated parameters that maximize expected returns are given by
\begin{eqnarray}
	&\hat{\Lambda} = V(:,1),    \\
	&\hat{\Phi} = U(:,1).
\end{eqnarray}
\end{proposition}
These estimates correspond to the first singular vectors from the matrices \(V\) and \(U\), respectively. This choice ensures that the optimal trading strategy leverages the directions that capture the greatest variance in the prediction matrix \(\Pi\), thereby extracting the most informative signal structure. Importantly, \(\hat{\Lambda}\) and \(\hat{\Phi}\) are obtained from the singular value decomposition of the sample-based matrix \(\Pi\), and should therefore be interpreted as empirical estimators rather than population parameters.

    \subsection{Maximizing Sharpe Ratio}
    \label{sec:max-Sharpe}

    % % % % % % % % % % % % % % % % % % 
    % % Propostion 4. % % % % % % % % % 
    % % % % % % % % % % % % % % % % % % 
    
The following propositions formulate the estimates that maximize the squared Sharpe ratio. Appendix \ref{apx:max_sharpe_ratio} provides the proof and detailed derivations.

	\begin{proposition}
		\label{prp:max_sharpe_ratio}

        Assume that $\Phi$ is given. Based on \eqref{eqn:SR}, define
        \begin{equation}
        	C_{\Phi} = B_{\Phi}^{-1} A_{\Phi}.
        \end{equation}
        The optimal $\Lambda$ is the principal eigenvector $\Lambda_{\max}$ of the eigenvalue problem
        \begin{equation}
        		\label{eqn:prop_SR_Phi}
        	C_{\Phi}\,\Lambda = \lambda\,\Lambda.
        \end{equation}

        Similarly, assume $\Lambda$ is given. Based on Equation \eqref{eqn:SR_alternative}, define $C_{\Lambda} = B^{-1}_{\Lambda} A_{\Lambda}$.  The optimal value for $\Phi$ is the largest eigenvector $\Phi_{max}$ of the following eigenvalue problem:        
        \begin{equation}
    		\label{eqn:prop_SR_Lambda}
    		C_{\Lambda} \Phi = \lambda \Phi.
    	\end{equation}

The optimal solutions for $\Lambda$ and $\Phi$ are obtained by iteratively applying these two equations until convergence. We further rescale each solution to have unit norm.
        
    \end{proposition}

 In this way, we utilize both alternative expressions for the Sharpe ratio in  Proposition \ref{prp:linear_metrics} to iteratively estimate the optimal parameters $\Lambda$ and $\Phi$. However, the eigenvalue problems in \eqref{eqn:prop_SR_Phi} and \eqref{eqn:prop_SR_Lambda} require computing the inverse of large matrices, which is challenging in high-dimensional settings. To address this, we propose the following proposition to iteratively estimate $\Lambda$ and $\Psi$.

    \begin{proposition}
	\label{prp:max_sharpe_ratio_ridge}
	
	Consider a set of managed-portfolios $\chi_{\Phi}$ of dimension $T \times M$:
    \begin{equation}
        \label{eqn:managed_portfolio_for_Lambda}
        \chi_\Phi = 
    \begin{bmatrix}
    	(\chi_\Phi)_{2}' \\
    	(\chi_\Phi)_{3}' \\
    	\vdots \\
    	(\chi_\Phi)_{T+1}' \\
    \end{bmatrix},
    \end{equation}
    where 
    \begin{equation}
    (\chi_{\Phi})_{s} = \Pi_s' \Phi. 
    \end{equation}
    
The problem in \eqref{eqn:prop_SR_Phi} is essentially an asset‐allocation exercise: it seeks to maximize the squared Sharpe ratio by investing in $\chi_\Phi$ composed of $M$ assets.
This is equivalent to estimating $\Lambda$ as the mean-variance efficient portfolio weights.
	
	Following \citet{britten1999sampling}, the estimate of $\Lambda$ is obtained from the following regression:
	\begin{equation}
		\label{eqn:hat_Lambda_ols}		
		\mathbf{1} = \chi_\Phi \Lambda  + \mathbf{u},
	\end{equation}
	where $\mathbf{1}$ is a $T$-vector of ones and $T$ denotes the sample size.
	To handle high-dimensional settings, we adopt ridge regression \citep{kelly2023FML, shen2024can}.
	The estimator for $\Lambda$ is then given by:
	\begin{equation}
		\label{eqn:hat_Lambda_ridge}		
		\hat \Lambda  = (\chi_\Phi' \chi_\Phi + \lambda I_M)^{-1} \chi_\Phi' \mathbf{1},
	\end{equation}
	where $\lambda$ is a Ridge-type parameter that shrinks the regression coefficients towards zero.
	
	Similarly, we define a set of managed-portfolios $\chi_{\Lambda}$ of dimension $T \times N^2$:
    \begin{equation}
        \label{eqn:managed_portfolio_for_Phi}
        \chi_\Lambda = 
    \begin{bmatrix}
    	(\chi_\Lambda)_{2}' \\
    	(\chi_\Lambda)_{3}' \\
    	\vdots \\
    	(\chi_\Lambda)_{T+1}' \\
    \end{bmatrix},
    \end{equation}
    where 
\begin{equation}
    (\chi_\Lambda)_{s} = \Pi_s \Lambda.
  \end{equation}
	The problem in \eqref{eqn:prop_SR_Lambda} is another asset allocation exercise: it seeks to maximize the squared Sharpe ratio by investing in $\chi_\Lambda$.
	The estimator for $\Phi$ is
	\begin{equation}
		\label{eqn:hat_Phi_ridge}		
		\hat \Phi  = (\chi_\Lambda' \chi_\Lambda + \lambda I_{N^2})^{-1} \chi_\Lambda' \mathbf{1}.
	\end{equation}

    \end{proposition}

 We highlight several key aspects of Sharpe ratio-maximization.
 
 First, the preceding propositions recast the problem as a managed-portfolio optimization, yielding the optimal weights for the tangency portfolio—or equivalently, for the stochastic discount factor (SDF).
 
 Second, we impose a common ridge penalty $\lambda$ when estimating both $\Lambda$ and $\Phi$, thereby enforcing uniform shrinkage across all components. This shared regularization parameter simplifies exposition, enhances replicability, and mitigates overfitting in finite samples.
 We implement the five-fold cross-validation scheme to select $\lambda$ dynamically.
 
 Third, although the generalized eigenvalue solution provides a population-level characterization of the Sharpe ratio-maximizer, in practice we replace the unknown moment matrices with their sample analogues and apply the same ridge penalty. Rather than solving a generalized eigenvalue problem directly, we cast the estimation as a single ridge-penalized regression. This approach recovers the optimal SDF direction in finite samples, improves numerical stability by shrinking weights on weak or collinear signals, and avoids the computational burden of eigendecomposition. The resulting weight vector exactly coincides with the theoretical maximizer under the ridge-regularized sample formulation.
 
Fourth, the solution to the Sharpe ratio-maximizing strategy can be interpreted as a regularized linear combination of the principal components (PCs) of the matrix \( \Pi \), with both \( \Lambda \) and \( \Phi \) estimated via ridge regressions on projected versions of \( \Pi \). Unlike expected return-maximizing approaches that primarily load on the leading PCs, this strategy optimizes portfolio weights across the full spectrum of PCs. As a result, it captures predictive signals even in low-variance directions---consistent with the findings of \citet{kelly2025universal}---and achieves superior risk-adjusted performance.

Fifth, our methodology for estimating the stochastic discount factor offers a distinct contribution to recent advances that emphasize firm characteristics (e.g., \cite{kelly2019characteristics, lettau2020factors, chen2024deep, feng2024deep, didisheim2024apt, cong2025growing, liu2025genetic}). Unlike these approaches that treat assets independently, we incorporate structured cross-asset dependencies. This not only enhances empirical performance in out-of-sample tests but also yields a more interpretable economic narrative for how information propagates across securities.

Cross-asset dependencies are also central to recent transformer-based approaches in asset pricing, which leverage multi-headed attention mechanisms to extract and aggregate predictive signals across assets. For instance, \citet{cong2022alphaportfolio} introduce AlphaPortfolio, a deep reinforcement learning framework with cross-asset attention networks (CAAN) that model interdependencies among securities. Similarly, the AIPM framework of \citet{Kelly2024aipm} embeds transformer architectures within the SDF, showing that nonlinear information sharing across assets can significantly improve empirical performance.

While these transformer models offer substantial modeling flexibility, our framework provides a complementary linear alternative that emphasizes transparency and interpretability. We capture cross-asset spillovers through a connection matrix $\Psi$, where each element $\Psi_{i,j}$ quantifies the predictive influence of asset $i$’s signals on asset $j$’s returns. Although related to the linear surrogate of the transformer, our approach differs in a key respect. The linear transformer models the attention matrix as a function of asset-level signals, requiring estimation of $O(M^3)$ parameters, where $M$ denotes the number of characteristics. In such setups, signal relevance and cross-asset dependencies are entangled within the signal space.

By contrast, we disentangle these two components: signal relevance is captured by a vector $\Lambda$, while cross-asset connections are modeled separately via $\Psi$. This separation reduces parameter complexity to $O(M + N^2)$, promotes computational efficiency, enables straightforward replication, and delivers an economically interpretable decomposition of predictive strength and cross-asset signal spillovers.

% \clearpage

\section{Data}
\label{sec:data}

Our dataset combines monthly stock returns from the Center for Research in Security Prices (CRSP), accounting variables from Compustat, and analyst coverage and earnings forecasts from the Institutional Brokers' Estimate System (IBES). We assume that quarterly and annual financial statements from Compustat become publicly available four months after the end of the corresponding fiscal quarter. The full sample spans January 1963 to December 2023. Out-of-sample evaluation begins in February 1973, with estimation windows based on rolling samples of the most recent 120 monthly observations.

\subsection{Predictive Characteristics }
\label{sec:characteristic_data}

We employ 138 firm-level signals across 13 characteristic themes: Accruals, Debt Issuance, Investment, Leverage, Low Risk, Momentum, Profit Growth, Profitability, Quality, Seasonality, Size, Short-Term Reversal, and Value. These signals originate from \citet{jensen2023there}.\footnote{We use the ``Global Stock Returns and Characteristics'' dataset under ``Contributed Data Forms'' on WRDS: \url{https://wrds-www.wharton.upenn.edu/pages/get-data/contributed-data-forms/global-factor-data/}. Table IA.II of \citet{jensen2023there} details the signal definitions and references. Of the original 153 signals, we exclude 15 that begin after 1963 to satisfy the sample-coverage requirements of \citet{kelly2023principal}. We apply standard filters to retain only observations with: (i) \texttt{excntry} = ``USA'', (ii) CRSP \texttt{shrcd} $\in\{10,11\}$, (iii) CRSP \texttt{exchcd} $\in\{1,2,3\}$, and (iv) non-missing monthly excess return (\texttt{ret\_exc}) and next-month excess return (\texttt{ret\_exc\_lead1m}). Each characteristic is standardized to have a mean of zero and a standard deviation of one.}

\subsection{Spread Portfolios}
\label{sec:spread_data}

For each of the 138 signals, we sort stocks into terciles each month and compute high-minus-low factor returns. To form factor-level signals, we aggregate stock-level signals into corresponding factor portfolios. Returns and signals are value-weighted by market equity, with individual market-equity weights winsorized at the 80th percentile of NYSE capitalization, following the data providers' recommendations.

\subsection{Bivariate Sorting on Size and Other Characteristics}
\label{sec:bisort_data}

We also construct bivariate sorted portfolios to serve as alternative investment universe. First, stocks are sorted into two size groups (big vs.\ small) based on market equity. Independently, each signal sorts stocks into three groups (high, medium, low). Cross-classifying these sorts produces six portfolios; we retain only the high and low portfolios for each size group, resulting in four portfolios per signal. As with the spread portfolios, returns and signals are capped-value-weighted by winsorized market equity. We omit the bivariate portfolios for the characteristic \texttt{ami\_126d} due to missing returns in 2023. Moreover, since size already plays a role in the sorting procedure, we consider a total of $136 \times 4 = 544$ portfolios.

Thus, we consider two investment universes: one constructed from univariate sorts comprising 138 spread portfolios, and the other from bivariate sorts comprising 544 portfolios. Each portfolio is associated with a time series of returns and 138 signal observations.

    \section{Results}
    \label{sec:results}
    
    \subsection{An Illustrative Toy Example}
    
   To build intuition for the proposed framework, we construct a low-dimensional toy dataset comprising five firm characteristics—market equity (ME), book-to-market ratio (BM), operating profits to lagged book equity (OP), asset growth (INV), and 12-month momentum (MOM)—and nine portfolios formed by a $3 \times 3$ sort on ME and BM (ranging from ME1×BM1 to ME3×BM3). 
   
 This simplified setup allows us to explicitly report the estimated low-dimensional parameters $\Lambda$ and $\Psi$, as well as the weight vector $\omega$. It also enables a comparison of key performance metrics for: (i) strategies subject to unit-norm constraints without an explicit zero-cost requirement; and (ii) zero-cost strategies with total leverage constrained to two.
    
    We implement expected return and Sharpe ratio-maximizing strategies, as formulated in sections \ref{sec:model} and \ref{sec:estimation}. These strategies, which target different objectives, yield notable differences in parameter estimates and performance outcomes. Table~\ref{tab:toy_performance} summarize the monthly average returns, monthly standard deviations, and \emph{annualized} Sharpe ratios for each strategy over the out‐of‐sample period from February~1973 to December~2023.
    
   The first two rows of the table consider the case in which the zero-cost assumption is not imposed. The results show that the strategy maximizing expected return (\text{MR\,Cross}) delivers a high average monthly return of 5.56\%, but with substantial volatility (standard deviation of 61.9\%), yielding a Sharpe ratio of just 0.31.
   
    The Sharpe ratio-maximizing strategy (MS\,Cross) attains a mean return of 2.36\% and a much lower volatility (9.78 \%), yielding a Sharpe ratio of 0.84. Consequently, a mean--variance investor would find the Sharpe ratio‐maximizing strategy considerably more attractive, whereas an investor solely targeting expected returns would prefer the expected return‐maximizing strategy. Thus far, the out‐of‐sample performance aligns closely with the ex ante investment objectives.
    
    Next, we consider a strategy that maximizes the Sharpe ratio using self‐prediction to isolate the incremental contribution of cross‐predictive relations relative to self‐predictive relations. The key distinction between these two strategies lies in the structure of the connection matrix $\Psi$. Under cross‐prediction, $\Psi$ is a full $9\times9$ matrix, capturing all pairwise interactions among the characteristics and returns of assets. In contrast, under self‐prediction, $\Psi$ is restricted to its diagonal terms.
    
   The second and third rows of Table~\ref{tab:toy_performance} report the performance of the Sharpe ratio–maximizing strategies under cross‐prediction and self‐prediction, respectively. The cross‐prediction strategy (MS\,Cross) delivers a Sharpe ratio of 0.84 with a mean return of 2.36\%, whereas the self‐prediction variant (MS\,Self) achieves a lower Sharpe ratio of 0.60 and the lowest mean return of 1.31\% . This gap in both risk‐adjusted and absolute returns illustrates the incremental benefit of incorporating cross‐predictive relationships beyond self‐prediction alone, underscoring the pivotal role of cross‐predictive dynamics in enhancing portfolio performance.
   
To provide further economic perspective on the value of accounting for cross-stock predictability, we compute the certainty equivalent return of the investment strategies. The certainty equivalent is defined as $CE = \mu - \frac{\gamma}{2} \sigma^2$, where $\mu$ and $\sigma$ are the expected return and volatility of the strategy, respectively, and the risk aversion parameter $\gamma$ is set to 2. Accounting for cross-predictability, the certainty equivalent rate of return is approximately 16.84\% per year—8.00\% higher than that of self-predictability—indicating economically significant gains.

We next maximize expected return and Sharpe ratio under the zero‐cost and leverage‐two constraints. The fourth and fifth rows of Table~\ref{tab:toy_performance} report these constrained strategies, confirming that imposing the zero‐cost restriction reduces expected returns for both objectives. Nevertheless, even with zero cost and fixed leverage, the Sharpe ratio–maximizing strategy outperforms the expected‐return–maximizing strategy, delivering a higher mean return (0.50\% vs.\ 0.49\%) and a substantially higher Sharpe ratio (1.22 vs.\ 0.53).

To provide additional insight into cross-prediction and self-prediction strategies, Table~\ref{tab:toy_lambda_psi} reports the estimated values of $\Lambda$, $\Psi$, and $\omega$ for each approach without imposing the zero-cost constraint. The estimation window spans 120 months, from December 2003 to November 2023, covering our last out-of-sample period. Panel~A presents the Sharpe ratio-maximizing cross-prediction strategy; Panel~B presents the Sharpe ratio-maximizing self-prediction strategy; and Panel~C reports the differences in the portfolio weights $\omega$ between the two.

In Table~\ref{tab:toy_lambda_psi}, Panel A shows that the estimated $\Lambda$ coefficient for book‐to‐market equity (BM) is $-0.34$, whereas the coefficients for the other four characteristics are all positive, with the smallest value at $0.21$. This suggests that the Sharpe ratio–maximizing strategy with cross‑prediction is well balanced across the five characteristics. The full $9\times9$ matrix $\Psi$ exhibits substantial values both on and off the diagonal: the average absolute value of its diagonal entries is $0.0068$, compared to an average absolute off‑diagonal entry of $0.0805$, indicating that cross‑predictive relationships play an even more substantial role in defining the trading strategy.
    
  Panel~B of Table~\ref{tab:toy_lambda_psi} shows that under self‐prediction the estimated $\Lambda$ coefficients exhibit greater dispersion---asset growth (INV) even turns negative---while $\Psi$ is constrained to its diagonal (average absolute value of $0.0288$, all off‐diagonals zero). 
  This contrast highlights the structural effect of cross-predictability: including cross-predictive terms not only yields nonzero off-diagonal elements of $\Psi$ but also shifts the estimated $\Lambda$ coefficients, altering the relative importance of characteristics.

  Panel~C reports how the optimal weights $\omega$ shift between cross‐ and self‐prediction: 
  under cross‐prediction, long exposures to ME3\,BM1 decrease, 
  and shorts in ME1\,BM2 deepen.
  For example, the ME1\,BM2 position is $-0.27$ under cross‐prediction---driven by off‐diagonal $\Psi$ entries of $0.24$ and $0.25$---whereas it is substantially smaller under self‐prediction. 
  
  As noted earlier, the optimal trading strategy that accounts for cross-predictability delivers a 8.00\% higher certainty equivalent return, suggesting that the estimated $\Lambda$ and $\Psi$, which determine the portfolio weights $\omega$, differ to an economically significant degree when cross-predictability is incorporated, relative to the benchmark case of self-predictability.

    In summary, the results in Tables~\ref{tab:toy_performance} and~\ref{tab:toy_lambda_psi} confirm that incorporating cross‑predictive relationships is valuable for constructing robust investment strategies, even in a low‑dimensional illustrative setting.

    % \clearpage
    
    % Table generated by Excel2LaTeX from sheet 'performance'
    \begin{table}[htbp] 
    \caption{Performance of Strategies of a Toy Example}
    \label{tab:toy_performance}
    
    {\footnotesize
   
    This table reports the monthly average return (\%), monthly standard deviation (\%), and annualized Sharpe ratio for five strategies in a low‐dimensional toy example involving five characteristics and nine assets. The strategies are:
    \begin{enumerate}
    	\item Unconstrained expected return-maximization with cross‐prediction;
    	\item Unconstrained Sharpe ratio-maximization with cross‐prediction;
    	\item Unconstrained Sharpe ratio-maximization with self‐prediction;
    	\item Zero‐cost, leverage‐two expected return-maximization with cross‐prediction;
    	\item Zero‐cost, leverage‐two Sharpe ratio-maximization with cross‐prediction.
    \end{enumerate}

    \begin{center}
    
    \begin{tabular}{lcccc}
    \toprule
    \\
    & \multicolumn{1}{l}{$\mu$} & \multicolumn{1}{l}{$\sigma$} & \multicolumn{1}{l}{SR} & Cost \\
    \\
    \hline
    \\
    % MR Cross & 1.96  & 21.89 & 0.31  & Not Zero Cost \\
    % MS Cross & 0.77  & 1.92  & 1.38  & Not Zero Cost \\
    % \\
    % {\color{red} here.}
    % MS Self & 0.43  & 1.45  & 1.02  & Not Zero Cost \\
    % \\
    % MR Cross ZC & 0.49  & 3.22  & 0.53  & Zero Cost \\
    % MS Cross ZC & 0.54  & 1.47  & 1.26  & Zero Cost \\
    % \\
    \\
    MR Cross    & 5.56  & 61.9  & 0.31 & Not Zero Cost \\
    MS Cross    & 2.36  & 9.78  & 0.84 & Not Zero Cost \\
    \\
    MS Self     & 1.31  & 7.57  & 0.60 & Not Zero Cost \\
    \\
    MR Cross ZC & 0.49   & 3.22  & 0.53 & Zero Cost \\
    MS Cross ZC & 0.50   & 1.43  & 1.22 & Zero Cost \\
    \\
    \bottomrule
    \end{tabular}%
    
    \end{center}
    
    }
    
    \end{table}%
    
    \clearpage
    % Table generated by Excel2LaTeX from sheet 'Sheet3'
    \begin{table}[htbp]
    \caption{Estimates for $\Lambda$ and $\Psi$ of a Toy Example}
    \label{tab:toy_lambda_psi}%
    {\footnotesize
    This tables reports the estimates for $\lambda$, $\Psi$, and $\omega$ of the Sharpe ratio-maximization strategies of the last rolling-window estimation, with cross-prediction in Panel A and self-prediction in Panel B. These strategies are free from zero-cost and leverage-two constraints.
    The $\Lambda$ vector has five elements corresponding to five characteristics: ME, BM, OP, INV, and MOM.
    There are nine assets for investment: the three-by-three sorted portfolios on ME and BM. 
    Specifically, they are ME1 BM1, ME1 BM2, ME1 BM3, ME2 BM1, ME2 BM2, ME2 BM3, ME3 BM1, ME3 BM2, ME3 BM3.
    The $\Psi$ is a nine-by-nine matrix, where the element i,j corresponds to the strength of the predictive relationship of the asset i's signals to asset j's returns.
    For cross-prediction in Panel A, the $\Psi$ has 81 values to estimate, while for self-prediction in Panel B, the $\Psi$ is only active in 9 values in the diagonal.
    In addition, the following two rows of panels A and B report the absolute average of the diagonal and off-diagonal terms of $\Psi$.
    Finally, Panel C shows the change of $\omega$ from cross- to self-prediction strategies.
    %}
    \vspace{-0.5cm}
    \begin{center}
    \resizebox{\textwidth}{!}{%
    \begin{tabular}{lccccccccc}
        \toprule
        \\
        \multicolumn{9}{c}{Panel A: Cross-Prediction} \\
        \\
        $\Lambda$ & ME    & BM    & OP    & INV   & MOM   &       &       &       &  \\
        & 0.21  & -0.34 & 0.29  & 0.53  & 0.69  &       &       &       &  \\
        \\
        $\Psi$  & 0.02  & -0.05 & 0.01  & -0.01 & 0.03  & -0.03 & -0.01 & -0.04 & 0.02 \\
          & 0.03  & -0.21 & -0.05 & 0.06  & 0.16  & -0.09 & 0.01  & -0.06 & 0.15 \\
          & 0.03  & 0.24  & -0.08 & -0.02 & -0.17 & 0.18  & -0.04 & 0.09  & -0.17 \\
          & -0.12 & 0.15  & 0.06  & 0.02  & 0.12  & -0.22 & 0.00  & -0.03 & 0.07 \\
          & 0.09  & -0.03 & -0.09 & -0.04 & -0.01 & 0.04  & 0.02  & 0.10  & -0.01 \\
          & -0.03 & -0.13 & 0.12  & -0.05 & -0.11 & 0.04  & 0.11  & -0.07 & 0.07 \\
          & 0.13  & -0.14 & 0.14  & 0.01  & -0.26 & 0.23  & 0.08  & 0.00  & -0.21 \\
          & -0.14 & 0.25  & -0.26 & -0.05 & 0.13  & 0.00  & -0.10 & 0.03  & 0.15 \\
          & -0.02 & -0.09 & 0.15  & 0.09  & 0.09  & -0.14 & -0.06 & -0.02 & -0.06 \\
        \\
        & \multicolumn{8}{l}{Absolute Average of Diagonal Terms $\Psi$} & 0.0068 \\
        & \multicolumn{8}{l}{Absolute Average of Off-Diagonal Terms $\Psi$} & 0.0805 \\
        \\
        $\omega$ & -0.03 & -0.27 & 0.20  & 0.00  & -0.03 & 0.03  & 0.09  & -0.06 & -0.06 \\
        % \\
        \hline
        \\
        \multicolumn{10}{c}{Panel B: Self-Prediction} \\
        \\
        $\Lambda$ & ME    & BM    & OP    & INV   & MOM   &       &       &       &  \\
        & -0.09 & -0.89 & 0.41  & -0.14 & 0.12  &       &       &       &  \\
        \\
        $\Psi$   & -0.14  & 0     & 0     & 0     & 0     & 0     & 0     & 0     & 0 \\
        & 0     & 0.51  & 0     & 0     & 0     & 0     & 0     & 0     & 0 \\
        & 0     & 0     & -0.13 & 0     & 0     & 0     & 0     & 0     & 0 \\
        & 0     & 0     & 0     & -0.04  & 0     & 0     & 0     & 0     & 0 \\
        & 0     & 0     & 0     & 0     & 0.41  & 0     & 0     & 0     & 0 \\
        & 0     & 0     & 0     & 0     & 0     & 0.02 & 0     & 0     & 0 \\
        & 0     & 0     & 0     & 0     & 0     & 0     & 0.35  & 0     & 0 \\
        & 0     & 0     & 0     & 0     & 0     & 0     & 0     & -0.63 & 0 \\
        & 0     & 0     & 0     & 0     & 0     & 0     & 0     & 0     & -0.09 \\
        \\
        & \multicolumn{8}{l}{Absolute Average of Diagonal Terms $\Psi$} & 0.0288 \\
        & \multicolumn{8}{l}{Absolute Average of Off-Diagonal Terms $\Psi$} & 0 \\    
        \\
        $\omega$ & -0.14 & -0.10 & 0.19  & -0.03 & 0.00  & -0.02 & 0.47  & -0.32 & 0.08 \\
        % \\
        \hline
        \\
        \multicolumn{10}{c}{Panel C: Change of Weights from Cross- to Self-Prediction} \\
        \\
        ID & ME1 BM1 & ME1 BM2 & ME1 BM3 & ME2 BM1 & ME2 BM2 & ME2 BM3 & ME3 BM1 & ME3 BM2 & ME3 BM3 \\
        \\
        & 0.10  & -0.17 & 0.00  & 0.03  & -0.03 & 0.05  & -0.37 & 0.26  & -0.13 \\
        % \\
        \bottomrule
    \end{tabular}%
    }
    \end{center}

    }
    
    \end{table}%
    
    \clearpage

    \begin{table}[h!]
    \caption{Performance of Cross-Predictive Strategies}
    \label{tab:linear}

    {\footnotesize
  
    This table reports the monthly average return (\%), monthly standard deviation (\%), and annualized Sharpe ratio of cross-predictive strategies. The strategies are zero-cost and leverage two. 
    MR and MS are strategies to maximize expected return and Sharpe ratio, respectively. 
    Panels A and C are for investing in 138 spread portfolios, and Panels B and D are for 544 bivariate sorted portfolios.
    In Panels A and B, we report the results of the whole out-of-sample period from February 1973 to December 2023 and the high and low sentiment periods split by the  sentiment median value over the sample periods from February 1973 to December 2023.
    In Panels C and D, we report for January 1990 to December 2023, and the high and low VIX periods split by the VIX median value over the sample periods from 1990 to 2023.
    
    \begin{center}
    
    \begin{tabular}{lccccccccccc}
    \toprule
    \\
    & \multicolumn{3}{c}{1973:02-2023:12} &       & \multicolumn{3}{c}{SENT High} &       & \multicolumn{3}{c}{SENT Low} \\
    \cline{2-4} \cline{6-8} \cline{10-12} 
    & $\mu$ & $\sigma$ & SR    &       & $\mu$ & $\sigma$ & SR    &       & $\mu$ & $\sigma$ & SR \\
    \\
    \hline
    \\
    \multicolumn{12}{c}{Panel A: Spread Portfolios} \\
    \\
    MR    & 0.42  & 3.23  & 0.45  &       & 0.73  & 3.79  & 0.67  &       & 0.11  & 2.53  & 0.15 \\
    MS    & 0.29  & 0.45  & 2.21  &       & 0.30  & 0.47  & 2.19  &       & 0.27  & 0.43  & 2.22 \\
    \\
    \hline
    \\
    \multicolumn{12}{c}{Panel B: BiSort Portfolios} \\
    \\
    MR    & 0.45  & 3.02  & 0.52  &       & 0.48  & 3.35  & 0.49  &       & 0.42  & 2.66  & 0.54 \\
    MS    & 0.26  & 0.27  & 3.32  &       & 0.28  & 0.27  & 3.58  &       & 0.24  & 0.27  & 3.08 \\
    \\
    \hline
    \hline
    \\
    & \multicolumn{3}{c}{1990:01-2023:12} &       & \multicolumn{3}{c}{VIX High} &       & \multicolumn{3}{c}{VIX Low} \\
    \cline{2-4} \cline{6-8} \cline{10-12} 
    & $\mu$ & $\sigma$ & SR    &       & $\mu$ & $\sigma$ & SR    &       & $\mu$ & $\sigma$ & SR \\
    \\
    \hline
    \\
    \multicolumn{12}{c}{Panel C: Spread Portfolios} \\
    \\
    MR    & 0.33  & 3.83  & 0.30  &       & 0.59  & 4.97  & 0.41  &       & 0.07  & 2.14  & 0.12 \\
    MS    & 0.24  & 0.43  & 1.92  &       & 0.30  & 0.52  & 2.02  &       & 0.18  & 0.31  & 1.98 \\
    \\
    \hline
    \\
    \multicolumn{12}{c}{Panel D: BiSort Portfolios} \\
    \\
    MR    & 0.39  & 3.20  & 0.42  &       & 0.57  & 3.87  & 0.51  &       & 0.20  & 2.33  & 0.30 \\
    MS    & 0.24  & 0.29  & 2.90  &       & 0.28  & 0.34  & 2.89  &       & 0.20  & 0.22  & 3.13 \\
    \\
    \bottomrule
    \end{tabular}%
        
    \end{center}
    
    }
    
    \end{table}

    \clearpage
    
\subsection{Zero-Cost Linear Strategies}

Table~\ref{tab:linear} reports the performance of linear cross‑predictive strategies implemented as zero‑cost, leverage‑two portfolios, comparable to common factor and anomaly implementations. MR and MS denote the strategies that maximize expected return and the Sharpe ratio, respectively. 
In Panel~A, we consider an investment universe with 138 spread portfolios detailed in the data section. Over the full sample period, MR achieves a monthly average return of 0.42\% with an annualized Sharpe ratio of 0.45, whereas MS records a lower monthly average return of 0.29\% but a substantially higher annualized Sharpe ratio of 2.21.

We further analyze performance during evolving market states by splitting the out‑of‑sample period at the median of the investor sentiment index \citep{baker2006investor}.\footnote{The sentiment data spans July 1965 to December 2023 and is obtained from the variable `SENT'' on Jeffrey Wurgler's website: \url{https://pages.stern.nyu.edu/~jwurgler/data/SENTIMENT.xlsx}.} During high‑sentiment regimes, MR delivers an average monthly return of 0.73\%, while in low‑sentiment regimes its return falls to 0.11\%. The MS strategy exhibits robust Sharpe ratios across both regimes: 2.19 in high‑sentiment periods and 2.22 in low‑sentiment periods.

In Panel~B, we evaluate investments in 544 bi‑variate sorted portfolios as detailed in the data section. Over the full out‑of‑sample period (January 1973–December 2023), MR delivers a monthly average return of 0.45\% and an annualized Sharpe ratio of 0.52, whereas MS achieves an exceptionally high annualized Sharpe ratio of 3.32. In sub‑period analyses, MR’s average return increases during high‑sentiment regimes, while MS maintains Sharpe ratios above 3 in both high‑ and low‑sentiment periods.

In Panels~C and~D, we split the period January 1990–December 2023 at the median of the VIX index.\footnote{The VIX data spans 1990 to 2023 and is obtained from the CBOE: \url{http://www.cboe.com/products/vix-index-volatility/vix-options-and-futures/vix-index/vix-historical-data/}.} In Panel~C (spread portfolios), MR's average return is 0.59\% during high‑VIX regimes and 0.07\% during low‑VIX regimes (0.33\% full sample), while MS records Sharpe ratios of 2.02 and 1.98 in high‑ and low‑VIX regimes (1.92 full sample).

In Panel~D (bi‑variate sorted portfolios), MR attains a monthly average return of 0.39\% and an annualized Sharpe ratio of 0.42, while MS achieves a Sharpe ratio of 2.90. MR’s return remains higher in high‑VIX regimes, and MS sustains Sharpe ratios around 3 in both high‑ and low‑VIX regimes.

In summary, MR strategies deliver high expected returns during high‑sentiment and high‑VIX regimes, but considerably lower expected returns otherwise. In contrast, MS strategies consistently achieve superior Sharpe ratios across all market states.

    % \clearpage

    \subsubsection{Comparing with Principal Portfolios (PP)}

We compare the principal portfolio-based trading strategies of \citet{kelly2023principal} with our own over the out-of-sample period from 1973 to 2019, as in their study. The results are reported in Table~\ref{tab:PP_MR_MS}.

Panel A, row 1 (PP--ME), reports the performance of the first principal portfolio on the market-equity signal: a 3.27\% monthly expected return, a 0.51 annualized Sharpe ratio, and a sum of absolute equity positions equal to 23.22. Rows 2 and 3 present the first principal portfolios for the book-to-market and momentum signals, which achieve Sharpe ratios of 0.60 and 0.48, respectively, with similarly high leverage. 
The principal portfolio can be applied to only one signal at a time. We also consider to take the 1/N equal-weighted strategy of the first principal portfolios across all 138 signals, namely the PPEW strategy, which delivers a 2.83\% monthly expected return and a 0.56 annualized Sharpe ratio. Notably, the leverage of PPEW is only 1.35, reflecting substantial diversification benefits by equal weighted average across predictors. 

Our maximum-expected return strategy  achieves an 135.14\% monthly expected return and a 0.52 annualized Sharpe ratio, with leverage of 537.70. Overall, the maximum-expected return strategy slightly underperforms the principal portfolios in Sharpe ratio, albeit remains reasonably close to them.

By contrast, the MS strategy harnesses multiple predictors to diversify exposures and optimize risk-adjusted returns, achieving an annualized Sharpe ratio of 2.22 with a leverage factor of 438.01. While the principal portfolio approach targets expected return subject to a volatility constraint, our strategy is derived directly from Sharpe ratio-maximization. As a result, it places greater emphasis on balancing return and risk, leading to improved performance on risk-adjusted metrics in our empirical setting.

To enhance implementability, we impose zero‐cost and leverage‐two constraints on both strategies. Panel~B of Table~\ref{tab:PP_MR_MS} reports the resulting performance. Under these constraints, the maximum‐expected‐return strategy (Row~1) achieves a 0.46\% monthly expected return and an annualized Sharpe ratio of 0.51, while the maximum‐Sharpe ratio strategy (Row~2) attains a 0.30\% monthly expected return and an annualized Sharpe ratio of 2.33. In both cases, the portfolios maintain zero net cost and a constant leverage of two in every period.

Overall, the maximum‑Sharpe ratio strategy remains highly competitive—delivering superior risk‑adjusted performance relative to a range of recent approaches, including principal portfolios. Accordingly, we focus our subsequent analyses to the constrained max‑SR strategy.

	% \clearpage
    
	\begin{table}[h!]
		\caption{A First Comparison on the Performance of PP, MR, and MS}
		\label{tab:PP_MR_MS}
		
		{\footnotesize
		
		This table reports each strategy’s monthly average return (\%), monthly standard deviation (\%), annualized Sharpe ratio, time-series average of the sum of positions on basic assets, and time-series average of the absolute sum of positions on basic assets. PP‑ME is the Principal Portfolio strategy using the market‑equity signal; PP‑BM uses book‑to‑market; PP‑MOM uses momentum. PP‑EW is an equal‑weighted combination of the first principal portfolios of 138 signals. MR is our maximum‑expected return strategy, and MS is our maximum‑Sharpe ratio strategy. Panel~A places no leverage or cost constraints on the strategies. Panel~B imposes two constraints—a zero‑cost requirement and a leverage restriction—on all strategies. Data span January~1963 through December~2019 (from PP's replication package on the \textit{Journal of Finance} website), with the out‑of‑sample period running from February~1973 to December~2019.

        \begin{center}
            
        \begin{tabular}{l ccccc}
            \toprule
            \\
            Strategy & $\mu$ & $\sigma$ & SR & Sum & ASum \\
            \\
            \hline
            \\
            \multicolumn{6}{c}{Panel A: Strategies} \\
            \\
            PP-ME    & 3.27  & 22.32 & 0.51  & 23.21 & 23.22 \\
            PP-BM    & 4.64  & 26.94 & 0.60  & 12.62 & 14.45 \\
            PP-MOM   & 3.65  & 26.41 & 0.48  & 23.94 & 25.34 \\
            PPEW     & 2.83  & 17.52 & 0.56  & 1.03  & 1.35 \\
            \\
            MR    & 135.14  & 895.59 & 0.52  & 95.83  & 537.70  \\
            MS    & 68.65   & 107.15   & 2.22  & 48.33   & 438.01 \\
            \\
            \hline
            \\
            \multicolumn{6}{c}{Panel B:  Strategies with Zero Cost} \\
            \\
            MR    & 0.46  & 3.10  & 0.51  & 0.00  & 2.00 \\
            MS    & 0.30  & 0.45  & 2.33  & 0.00  & 2.00 \\
            \\
            \bottomrule
        \end{tabular}%

        \end{center}
		}
	\end{table}

    \clearpage
    
    % Table generated by Excel2LaTeX from sheet 'Sheet1'
    \begin{table}[h!]
      \caption{Cross- vs Self-Prediction}
      \label{tab:self}
      {\footnotesize

    This table reports the monthly average returns (\%), monthly standard deviation(\%), and annualized Sharpe ratio, time-series average of the sum of positions on basic assets, and time-series average of the absolute sum of positions on basic assets. The objective of the strategies is to maximize the Sharpe ratio. 
    Notably, the cross-prediction strategies can be solved with and without the zero-cost constraint; however, the self-prediction strategy does not have an analytic solution once adding the zero-cost constraint.
    % , see Section \ref{sec:economic_restrictions} for details.
    Panel A invests on spread portfolios and Panel B is for bivariate sorted portfolios.
    The out-of-sample period is February 1973 to December 2023.

      \begin{center}
        \begin{tabular}{lccccc}
        \toprule
        \\
        & $\mu$ & $\sigma$ & SR & Sum & ASum \\
        \\
        
        \hline
         \\
         \multicolumn{6}{c}{Panel A: Spread Portfolios} \\
         \\
         MS Self & 6.76  & 16.54  & 1.42  & 1.26  & 30.93 \\
         \\
         MS Cross       & 64.43 & 107.19 & 2.08  & 45.06 & 438.46 \\ 
         MS Cross ZC    & 0.29  & 0.45   & 2.21  & 0.00  & 2.00 \\
         \\

        \hline
        \\
        \multicolumn{6}{c}{Panel B: BiSort Portfolios} \\
        \\
        MS Self & 15.83  & 26.64  & 2.06  & 2.34 & 99.05 \\
        \\
        MS Cross    & 389.99 & 450.02 & 3.00  & 112.14 & 2154.07 \\
        MS Cross ZC & 0.26   & 0.27   & 3.32  & 0.00    & 2.00 \\
        \\
        
        \bottomrule
        \end{tabular}%
        \end{center}
          
          }
          
    \end{table}%

	% \clearpage

    \subsubsection{Cross-Prediction SDF versus Self-Prediction SDF}
    \label{sec:cross-vs-self}
    
   The existing literature on SDF estimation has predominantly focused on self-predictive frameworks, where each asset’s signals are used solely to forecast its own returns.
    \cite{kelly2019characteristics} propose Instrumented PCA with the belief that the factor loadings on SDF are determined by assets characteristics, overcoming the limitations of static loading in PCA.
    \cite{lettau2020factors} find that the SDF estimated on Risk-Premium PCA is more highly correlated with the true SDF than those estimated on PCA.
    \cite{luo2025sdf} estimate the SDF with observable characteristics-based factors with L1-penalized SDF regression; whereas, \cite{didisheim2024apt} apply the L2-penalized SDF regression on observable and Random-Fourier-Feature generated factors. 
    All of these papers have been working on high-dimensional characteristics-based portfolios to estimate the SDF, where the belief of self-prediction are embedded the portfolios.
    
    By contrast, our framework utilizes managed-portfolios inherently reflecting the belief of cross-prediction: $\pi_s$ \eqref{eqn:expected_return}, $\chi_\Phi$ \eqref{eqn:managed_portfolio_for_Lambda}, and $\chi_\Lambda$ \eqref{eqn:managed_portfolio_for_Phi}.
    Whether cross-predictive strategies—where signals from one asset help predict the returns of others—can systematically outperform self-predictive ones in high-dimensional settings remains an open question. 
   To investigate this, we construct the self-predictive strategy by restricting the matrix~$\Psi$ to its diagonal, thereby eliminating all cross-asset interactions. 
      
   Panel~A of Table~\ref{tab:self} reports the empirical performance of the Sharpe ratio–maximizing strategies on the 138 spread portfolios. The self-predictive strategy achieves a Sharpe ratio of 1.42, while the cross-predictive counterparts attain Sharpe ratios of 2.08 without zero-cost requirement and 2.21 with zero-cost and leverage-two constraints. This more than 0.60 difference in Sharpe ratio underscores the incremental value of incorporating cross-asset predictive signals.
   
   Panel~B reports results for the 544 bivariate sorted portfolios. The self-predictive strategy achieves a Sharpe ratio of 2.06, while the cross-predictive variants reach 3.32 and 3.00 under constrained and unconstrained implementations, respectively. This gap of more than 1.00 in Sharpe ratio highlights the significant contribution of off-diagonal elements in~$\Psi$ to improved portfolio performance.
   
   Overall, the evidence confirms that cross-predictive strategies materially enhance the estimation and performance of stochastic discount factors—particularly in richer portfolio universes and longer samples.

    % \clearpage
    
\subsubsection{Factor Spanning Tests}

We conduct a series of factor-spanning tests to assess whether the established asset pricing factors fully explain the expected returns of the Sharpe ratio-maximizing strategies. 
Table~\ref{tab:factor_spanning_test} reports monthly alphas (\%), factor loadings, and associated $t$-statistics. 
Panel~A presents the Sharpe ratio-maximizing strategy on the spread portfolios, while Panel~B reports for the bivariate sorted portfolios.  

We first evaluate the \citet{fama2015five} five-factor model (FF5). 
The strategy on spread portfolio exhibits modest loadings on Market ($\beta = -0.01$, $t = -2.56$) and SMB ($\beta = 0.01$, $t = 1.78$) but insignificant exposures to the other four factors, while delivering a highly significant monthly alpha of 0.29\% ($t = 13.29$). 
This suggests that the strategy's returns are largely orthogonal to the FF5 factors.
Also, we augment the FF5 model with momentum (UMD), short-term reversal (REV), and liquidity (LIQ) factors \citep{pastor2003}. 
In this expanded specification, the strategy shows significant loadings on UMD and REV but not on LIQ, while its alpha remains economically and statistically significant at 0.26\% ($t = 11.54$). 
These findings indicate that momentum and reversal effects partially explain the strategy's performance, with little role for liquidity risk.

Next, we then examine the \citet{hou2015} four-factor model, which incorporates investment (R\_IA) and profitability (R\_ROE) factors alongside market and size factors. 
The strategy displays negligible loadings on R\_IA and R\_ROE, while maintaining a highly significant alpha.
The \citet{stambaugh2017} mispricing factors---MGMT and PERF---also fail to subsume the strategy's returns: 
The strategy shows minimal exposures to both factors, with an alpha of 0.28\% ($t = 10.20$).
Then, we assess the \citet{daniel2020short} model, which includes the market factor and two behavioral factors, PEAD and FIN.
While the strategy loads significantly on PEAD, its alpha remains robust at 0.29\% ($t = 11.09$), and it shows no meaningful exposure to FIN.
Finally, in a comprehensive regression incorporating all fourteen factors, The strategy maintains an alpha of 0.26\% ($t = 8.04$), with statistically significant but economically small loadings on SMB, UMD, REV, LIQ, FIN, and R\_IA. 
These results collectively demonstrate that the  strategy's expected returns cannot be fully explained by existing factor models.

Panel~B corroborates these findings. The strategy on bivariate sorted portfolio displays statistically significant but economically modest loadings on RMW, CMA, REV, PERF, R\_IA, MGMT, and PERF. 
Notably, it maintains a monthly alpha of 0.25\% ($t = 11.36$) even after controlling for all fourteen factors, further supporting the strategy's robustness to established factor models.

Across all specifications—including the Fama–French five-factor model with UMD, REV, and LIQ augmentations, Hou–Xue–Zhang, Stambaugh–Yuan, and Daniel–Hirshleifer–Sun frameworks, and even the comprehensive fourteen-factor regression—the Sharpe ratio-maximizing strategies on the spread  and bivariate sorted portfolios exhibit persistently large and highly significant alphas with only moderate loadings on existing factors. 
This suggests that conventional models may miss the cross-asset return predictability captured by our strategy.
Below, we further analyze the pricing content of the Sharpe ratio-maximizing strategies.

    \begin{table}
    \caption{Alpha and Factor Loadings}
    \label{tab:factor_spanning_test}%
    {\footnotesize
        This table reports the monthly alphas (\%), factor loadings, and their $t$-values (in parentheses) obtained from the factor-spanning tests of regressing the strategy returns on other asset pricing factors. 
        We have scaled the original strategy and factor returns by 100 for percentage compatibility, aiding coefficient comparability. This table focus on the Sharpe ratio-maximizing strategies with zero cost and leverage two.
        Panel A displays the results for investing in spread portfolios, while Panel B shows for bivariate sorted portfolios.
        The factors include FF5 factor, momentum factor (UMD), short-term reversal factor (REV), liquidity factor (LIQ) from \cite{pastor2003}, short-horizon inattention factor (PEAD) and long-horizon financing factor (FIN) from \cite{daniel2020short}, investment factor (R\_IA) and return on equity factor (R\_ROE) from \cite{hou2015}, management factor (MGMT) and performance factor (PERF) from \cite{stambaugh2017}. 
        PEAD and FIN are available before December 2018. 
        MGMT and PERF are available before December 2016. 
        All other factors are available during the whole sample period. 
        We report with the \cite{neweywest1987} $t$-statistics using a Bartlett kernel and lag length $L = 4(T/100)^{2/9}$.
        One, two, and three asterisks indicate significance at the 10\%, 5\%, and 1\% levels, respectively.
    
    \begin{center}
    \resizebox{\textwidth}{!}{
   
    \begin{tabular}{lllllllllllllll}
    \toprule
    \\
    Alpha & Market & SMB & HML & RMW & CMA & UMD & REV & LIQ & PEAD & FIN & R\_IA & R\_ROE & MGMT & PERF \\
    \\
    \hline
    \\
    \multicolumn{15}{c}{Panel A: Spread Portfolios} \\
    \\

    0.29*** & -0.01** & 0.01* & -0.01 & 0.01 & 0.02 &  &  &  &  &  &  &  &  &  \\
    (13.29) & (-2.56) & (1.78) & (-1.02) & (0.98) & (1.53) &  &  &  &  &  &  &  &  &  \\
    0.26*** & -0.01** & 0.01 & -0.00 & 0.01 & 0.01 & 0.03*** & 0.02*** & -0.00 &  &  &  &  &  &  \\
    (11.54) & (-2.26) & (1.48) & (-0.16) & (0.68) & (0.91) & (3.34) & (2.74) & (-0.82) &  &  &  &  &  &  \\
    0.29*** & -0.01 &  &  &  &  &  &  &  & 0.03** & 0.01 &  &  &  &  \\
    (11.09) & (-1.56) &  &  &  &  &  &  &  & (2.14) & (0.91) &  &  &  &  \\
    0.28*** & -0.01** & 0.01* &  &  &  &  &  &  &  &  & 0.01 & 0.01 &  &  \\
    (12.91) & (-2.56) & (1.95) &  &  &  &  &  &  &  &  & (0.73) & (1.24) &  &  \\
    0.28*** & -0.00 & 0.02** &  &  &  &  &  &  &  &  &  &  & 0.02 & 0.03*** \\
    (10.20) & (-0.47) & (2.51) &  &  &  &  &  &  &  &  &  &  & (1.19) & (3.66) \\
    0.26*** & -0.01 & 0.01 & -0.00 & 0.01 & -0.03 & 0.01* & 0.03*** & -0.00 & 0.01 & 0.00 & 0.04 & -0.05*** & 0.01 & 0.04*** \\
    (8.04) & (-0.92) & (0.83) & (-0.13) & (0.44) & (-0.72) & (1.74) & (3.05) & (-0.51) & (0.90) & (0.37) & (1.23) & (-2.92) & (0.51) & (2.61) \\
    
    \\
    \hline
    \\
    \multicolumn{15}{c}{Panel B: BiSort Portfolios} \\
    \\
    
    0.25*** & -0.00 & 0.01* & -0.00 & 0.03*** & 0.01 &  &  &  &  &  &  &  &  &  \\
    (16.67) & (-0.62) & (1.94) & (-0.62) & (3.96) & (1.20) &  &  &  &  &  &  &  &  &  \\
    0.24*** & -0.00 & 0.01 & 0.00 & 0.02*** & 0.01 & 0.02** & 0.01 & -0.00 &  &  &  &  &  &  \\
    (13.98) & (-0.01) & (1.59) & (0.72) & (3.92) & (0.63) & (2.21) & (1.09) & (-0.37) &  &  &  &  &  &  \\
    0.25*** & 0.00 &  &  &  &  &  &  &  & 0.02** & 0.02*** &  &  &  &  \\
    (13.15) & (1.23) &  &  &  &  &  &  &  & (1.99) & (2.70) &  &  &  &  \\
    0.24*** & -0.00 & 0.01** &  &  &  &  &  &  &  &  & 0.01 & 0.02*** &  &  \\
    (14.61) & (-0.53) & (2.03) &  &  &  &  &  &  &  &  & (1.23) & (2.68) &  &  \\
    0.26*** & 0.01* & 0.01 &  &  &  &  &  &  &  &  &  &  & 0.02*** & 0.02*** \\
    (14.12) & (1.81) & (1.55) &  &  &  &  &  &  &  &  &  &  & (3.02) & (3.92) \\
    0.25*** & 0.01 & 0.01 & 0.00 & 0.02* & -0.02 & 0.01 & 0.01 & 0.00 & 0.01 & 0.00 & 0.03 & -0.02 & 0.01 & 0.02*** \\
    (11.36) & (1.42) & (1.58) & (0.15) & (1.88) & (-0.96) & (1.04) & (1.51) & (0.12) & (1.38) & (0.39) & (1.11) & (-1.29) & (1.05) & (2.69) \\
    
    \\
    \bottomrule
    \end{tabular}

    }
    
    \end{center}

    }
    
    \end{table}

  \clearpage

    \subsubsection{Evolution of Sharpe Ratios over the Sample Period}

\begin{figure}[htbp]
        \caption{Sharpe Ratio of Strategies}
        \label{fig:rolling_Sharpe_ratio}

        {\footnotesize
        
        The figure depicts ten-year trailing (annualized) Sharpe ratios for the cross-predictive maximum-Sharpe ratio strategies.
        ``MS Spread'' is the strategy to maximize Sharpe ratio investing in the spread portfolios.``MS BiSort'' is the strategy to maximize Sharpe ratio investing in the bivariate sorted portfolios. 
        ``MS BiSort fixed'' uses a fixed $\lambda=1$, which is the most frequently selected value.
        Panel A shows the Sharpe ratio, while Panel B shows the Sharpe ratio of each strategy divided by that of ``MS BiSort.''
        The out-of-sample period is from February 1973 to December 2023 in monthly frequency, and the first ten-year Sharpe ratio is obtained for January 1983.
        For comparison, the market factor (MKT-RF) and momentum factor (UMD) are included.
        }

        \begin{center}

		\begin{subfigure}[b]{0.8\textwidth}
			\centering
			\includegraphics[width=\textwidth]{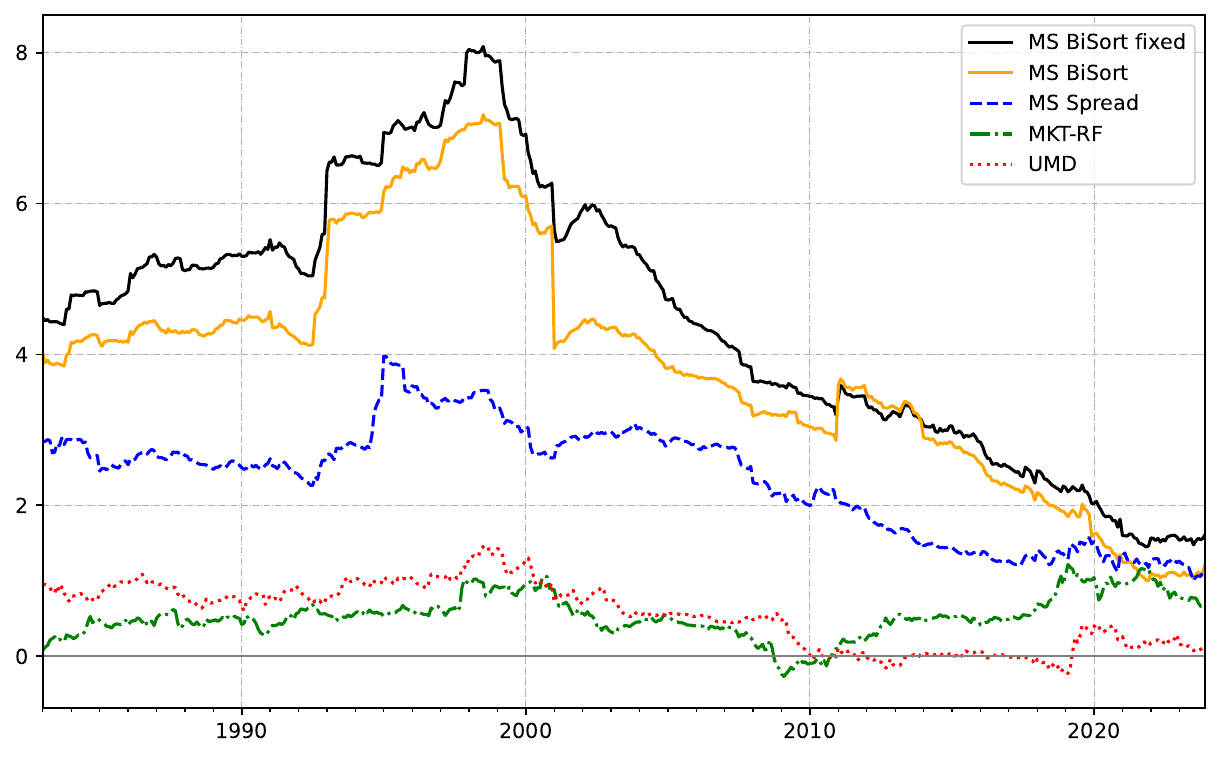}
			\caption{Sharpe Ratio}
		\end{subfigure}

		\begin{subfigure}[b]{0.8\textwidth}
			\centering
			\includegraphics[width=\textwidth]{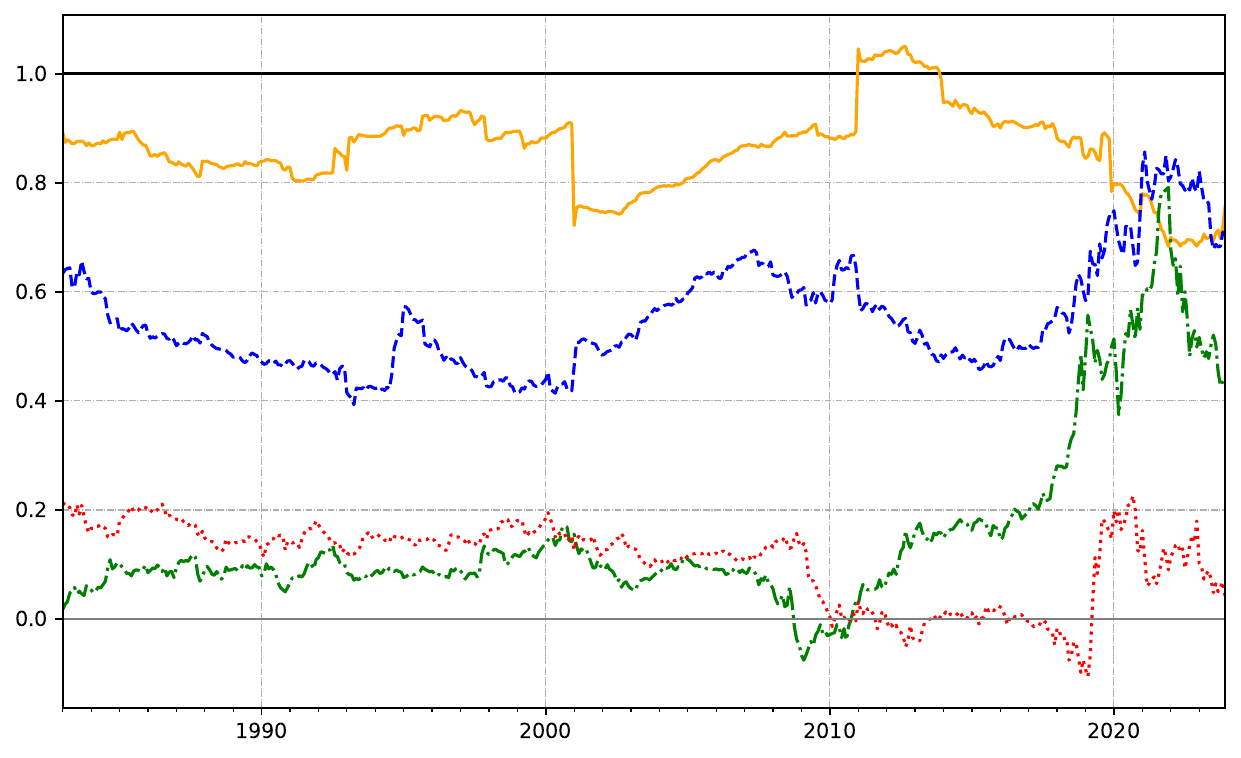}
			\caption{Sharpe Ratio Relative to MS BiSort}
		\end{subfigure}

        \end{center}

    \end{figure}

   To assess the persistence and evolution of risk‐adjusted returns over time, Figure~\ref{fig:rolling_Sharpe_ratio} Panel A shows the ten‐year trailing Sharpe ratios of our three maximum-Sharpe ratio strategies—MS Spread, MS BiSort and MS BiSort fixed—alongside those of the market and momentum factors for comparison. 
   \footnote{The shrinkage parameter $\lambda$ for MS Spread and MS BiSort are selected via cross-validation. 
   Appendix~\ref{sec:cross-validation} reports the selected parameter values of time.
   MS BiSort fixed uses a fixed $\lambda=1$, which is the most frequently selected value.
   }
   By smoothing over a decade window, we can observe how the trading strategies respond to changing market conditions.
   
   These strategies deliver eye‐catching Sharpe ratios in the 1990s—MS BiSort climbs as high as 4–7 before 2000, and MS Spread approaches 4—reflecting their ability to capture persistent value-enhancing opportunities. After 2000, however, it is natural to see some attenuation: wider adoption of anomaly tradings, increased market liquidity, and a lower‐volatility regime tend to compress excess returns over time. Accordingly, by the end of 2023, the trailing Sharpe of MS Spread and MS BiSort has moderated to about 1.2.
   By contrast, the MS BiSort fixed seems to deliver even higher Sharpe ratios than MS BiSort, suggesting that our cross-validation scheme is conservative and provides a low bound for the out-of-sample performance.
   
    To make more clear comparison, Figure~\ref{fig:rolling_Sharpe_ratio} Panel B shows the Sharpe ratio of each strategy relative to that of MS BiSort fixed. 
    In early sample before 2000, the Sharpe ratio of MS Spread (MS BiSort) is approximately 60\% (90\%) that of MS BiSort fixed, and market and momentum factors have below 20\% Shape ratio relative to MS BiSort.
    In the most recent sample, the Sharpe ratios of MS Spread, MS BiSort, MKT-RF, and UMD  are 70\%, 75\%, 43\%, and 4\% that of MS BiSort fixed.
 
    For context, both the market factor’s rolling Sharpe ratio and that of the momentum factor remain well below our strategies over the entire forty‐year span. Although the performance gap narrows in the post‐2000 era, both maximum-Sharpe ratio strategies continue to deliver robust risk‐adjusted returns relative to these benchmarks.

    Table \ref{tab:Sharpe_compare_FF} reports the (annualized) Sharpe ratios of the cross-predictive maximum-Sharpe ratio strategies, Fama-French five factors, and momentum factor for three sample periods: the whole OOS period from 1973:02 to 2023:12, before 2000:01, and after 2000:01. Although, the Sharpe ratios of our strategies attenuate after 2000, they remain competitive compared to the benchmark factors in three sample periods.

\clearpage

\clearpage

% Table generated by Excel2LaTeX from sheet 'out_Sharpe_Ratio'
\begin{table}[h!]
  \caption{Comparing Sharpe Ratios with Prevailing Factors}
  \label{tab:Sharpe_compare_FF}%
  
    {\footnotesize
    
    The table reports the (annualized) Sharpe ratios of the cross-predictive strategies, Fama-French five factors, and momentum. Three sample periods are 1973:02 to 2023:12, 1973:02 to 1999:12, and 2000 to 2023.
    
    \begin{center}
      
    \begin{tabular}{lcccccccc}
    \toprule
    \\
    & MS Spread & MS BiSort & MKT-RF & SMB   & HML   & RMW   & CMA   & UMD \\
    \\
    \hline
    \\

    1973-2023 & 2.21  & 3.32  & 0.45  & 0.21  & 0.33  & 0.45  & 0.5   & 0.45 \\
    1973-1999 & 2.84  & 4.98  & 0.48  & 0.16  & 0.47  & 0.36  & 0.58  & 0.96 \\
    2000-2023 & 1.58  & 2.21  & 0.41  & 0.27  & 0.20  & 0.54  & 0.43  & 0.09 \\
    
    \\
    \hline
    \end{tabular}%

    \end{center}

    }
  
\end{table}%

\begin{table}[h!]
\caption{Top Ten Signals by $\Lambda$}
\label{tab:signal_importance_top_ten}

{\footnotesize

    This table reports the top ten most important signals, Panel A for spread portfolios and Panel B for bivariate sorted portfolios. The columns are abbreviation, theme, time-series average of absolute $\Lambda$, full name, and original publication of the signals. There are 13 themes following \cite{jensen2023there}.    
    \\

    \resizebox{\textwidth}{!}{
    \begin{tabular}{llllll}
    \toprule
    \\
    & Abbreviation & Theme & $\Lambda$ & Full Name & Publication  \\
    \\
    \hline

    \\
    \multicolumn{6}{c}{Panel A: Spread Portfolio } \\
    \\

    2     & aliq\_at   & Investment & 0.139      & Liquidity of book assets                & \cite{ortiz2014real} \\
    34    & div12m\_me & Value      & 0.126      & Dividend yield                          & \cite{litzenberger1979effect} \\
    7     & at\_me     & Value      & 0.125      & Assets-to-market                        & \cite{eugene1992cross} \\
    9     & be\_gr1a   & Investment & 0.125      & Change in common equity                 & \cite{richardson2005accrual} \\
    44    & emp\_gr1   & Investment & 0.123      & Hiring rate                             & \cite{belo2014labor} \\
    45    & eq\_dur    & Value      & 0.123      & Equity duration                         & \cite{dechow2004implied} \\
    24    & col\_gr1a  & Investment & 0.123      & Change in current ope. lia. & \cite{richardson2005accrual} \\
    116   & sale\_gr3  & Investment & 0.123      & Sales growth (3 years)                  & \cite{lakonishok1994contrarian} \\
    15    & bev\_mev   & Value      & 0.121      & Book-to-market equity                   & \cite{penman2007book} \\   
    10    & be\_me     & Value      & 0.120      & Book-to-market enterprise value         & \cite{rosenberg1985persuasive} \\   
    \\

    \hline
    \\
    \multicolumn{6}{c}{Panel B: BiSort Portfolio } \\
    \\
    
    71    & ni\_be           & Profitability & 0.141      & Return on equity               & \cite{haugen1996commonality}              \\
    86    & ope\_bel1        & Profitability & 0.138      & Ope. profits-to-lagged be      & \cite{fama2015five}                \\
    85    & ope\_be          & Profitability & 0.136      & Operating profits-to-be        & \cite{fama2015five}                \\
    77    & o\_score         & Profitability & 0.135      & Ohlson O-score                 & \cite{dichev1998risk}                       \\
    42    & ebit\_sale       & Profitability & 0.134      & Profit margin                  &  \cite{soliman2008use}                      \\
    90    & prc              & Size          & 0.133      & Price per share                & \cite{miller1982dividends}            \\
    41    & ebit\_bev        & Profitability & 0.132      & Return on net operating assets &  \cite{soliman2008use}                  \\
    16    & bidaskhl\_21d    & Low Leverage  & 0.130      & The high-low bid-ask spread    & \cite{corwin2012simple}           \\
    65    & mispricing\_perf & Quality	     & 0.126      & Performance Based Mispricing   & \cite{stambaugh2017}  \\
    58    & ivol\_capm\_252d & Low Risk      & 0.126      & Idio. vol. to CAPM (21 days)   & \cite{ali2003arbitrage}       \\
    
    \\
    \bottomrule
    \end{tabular}
    }

}
    
\end{table}

\clearpage

\begin{figure}[htbp]
	\caption{Signal Importance}
	\label{fig:signal_importance}
	{\footnotesize
		
		This figure depicts the heatmaps of signal importance $\Lambda$ for each rolling-window estimation. 
        Sub-figures (a) and (b) are for spread portfolios and bivariate sorted portfolios, respectively.
        For interpretation, we focus on the absolute value of elements in $\Lambda$.
        We calculate the theme-level importance as the average of all signal-level importance within each theme.
        There are 13 themes following \cite{jensen2023there}.

	\begin{center}
		
		\begin{subfigure}[b]{0.9\textwidth}
			\centering
			\includegraphics[width=\textwidth]{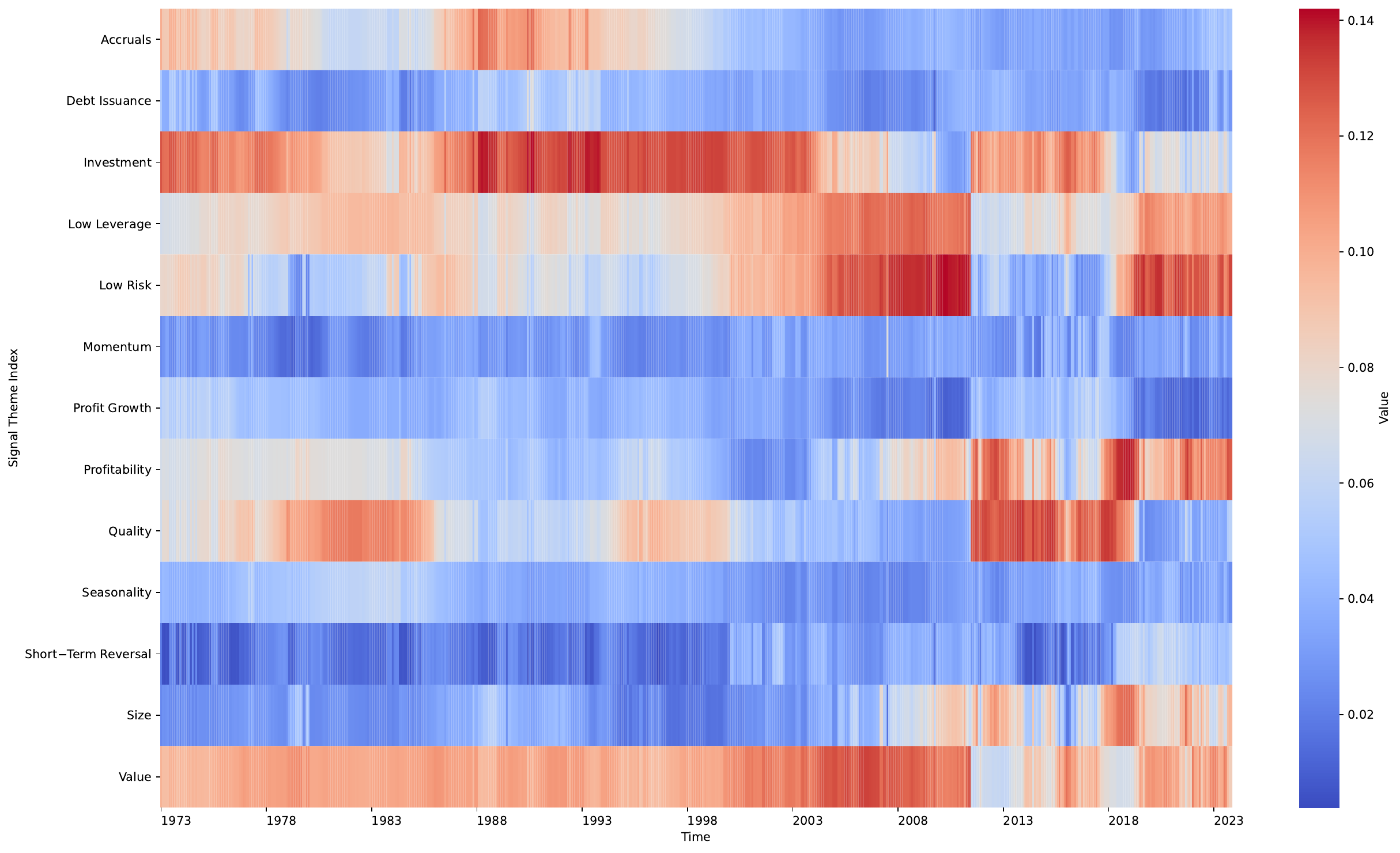}  
			\caption*{(a) Spread Portfolio: Theme-Level}
		\end{subfigure}

		\begin{subfigure}[b]{0.9\textwidth}
			\centering
			\includegraphics[width=\textwidth]{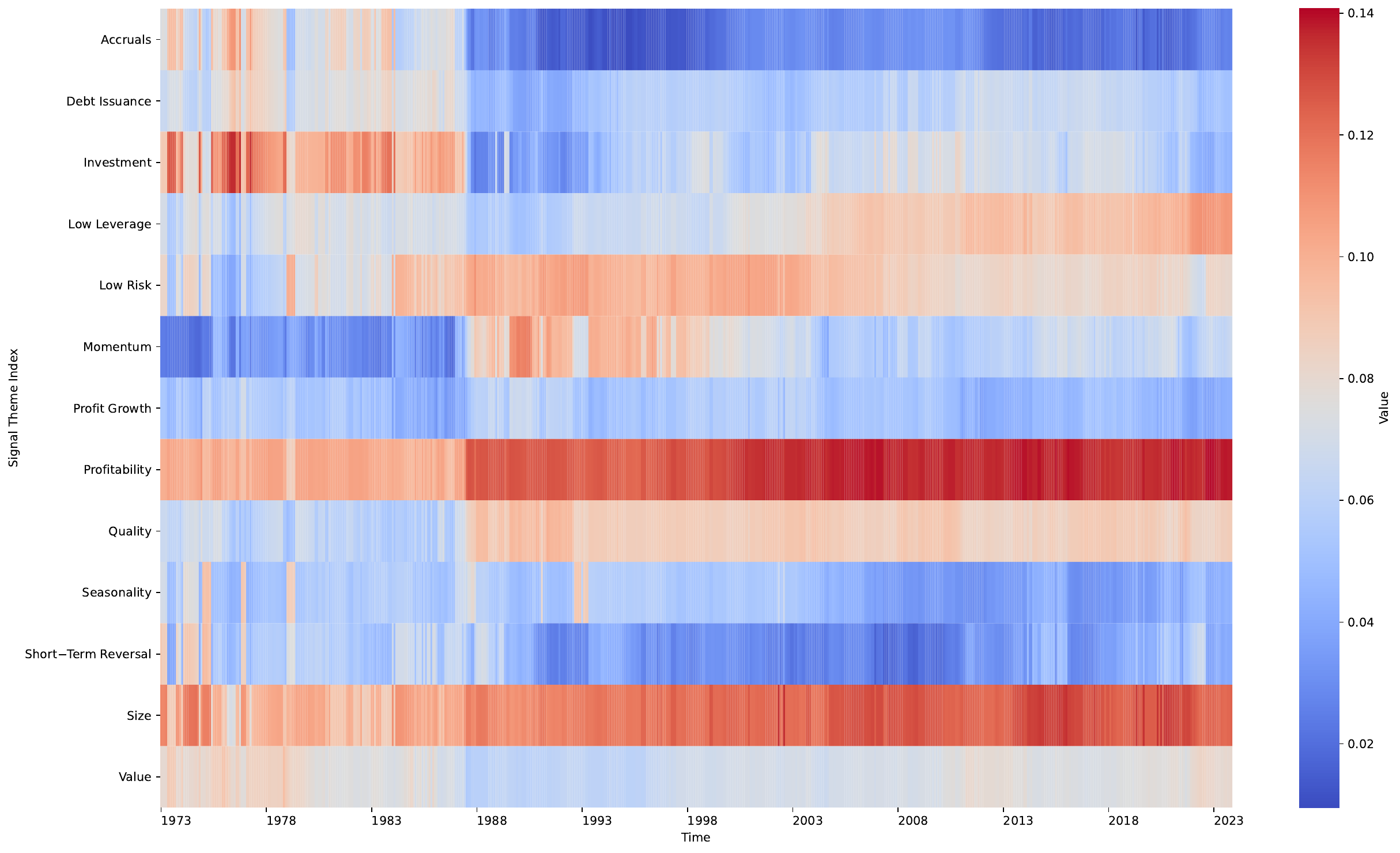}  
			\caption*{(b) BiSort Portfolio: Theme-Level}
		\end{subfigure}
        
	\end{center}

    }
	
\end{figure}

\subsection{Signal Importance}
\label{sec:emp_Lambda}

To understand the economic underpinnings of our Sharpe ratio-maximizing strategies or SDF, we examine the estimated values of $\Lambda$, which assign weights to firm-level predictive signals. These weights reflect the relative contribution of each signal to the SDF. We focus on the absolute value of these weights averaged over time to assess long-term signal importance.
Table~\ref{tab:signal_importance_top_ten} presents the ten most influential signals, ranked by their time-series average of absolute $|\Lambda|$ values, where Panel A is for spread portfolios and Panel B is for bivariate sorted portfolios.

Panel A, investing in spread portfolios, indicates that the most important signals are concentrated in the \emph{investment} and \emph{value} categories. For instance, the top signal---\emph{liquidity of book assets} \citep{ortiz2014real}---receives an average importance of 0.139, while \emph{dividend yield} \citep{litzenberger1979effect}, the leading signal in the value theme, ranks seventh overall with an importance of 0.126. These findings suggest that the strategy places greater emphasis on firm fundamentals linked to capital structure, financing constraints, and valuation, rather than technical or return-based indicators.

As for Panel B, \emph{profitability} dominates the top ten signals, followed by \emph{size}, \emph{low leverage}, and \emph{low risk} themes. For instance, \emph{return on equity} \citep{haugen1996commonality} and \emph{operating profitability-to-lagged book equity} \citep{fama2015five} are top signals, all belonging to \emph{profitability}.
Besides, \emph{price per share} \citep{miller1982dividends} emerges from the \emph{size} theme, recalling stronger size effects in the test assets sorted on size and other signals.

% \clearpage

Figure~\ref{fig:signal_importance} presents the importance measures for all 138 signals, organized into 13 thematic categories (as defined in the Data section). Sub-figures (a) and (b) display theme-level importance for spread portfolios and bivariate-sorted portfolios, respectively.\footnote{We provide the time-varying signal-level importance measures in Figure~\ref{fig:signal_importance_app} of the Appendix~\ref{sec:signal_importance_app}.} The heatmap visualization employs color intensity to indicate importance levels---with red (blue) representing high (low) importance--allowing clear identification of which signals consistently influence portfolio construction. 

In sub-figure (a) for spread portfolios, \emph{investment}- and \emph{value}-related signals dominate the red spectrum, reinforcing the role of tangible firm fundamentals. In contrast, \emph{momentum}, \emph{profit growth}, \emph{debt issuance}, \emph{seasonality}, and \emph{short-term reversal} appear consistently in the blue range, indicating minimal weight in the Sharpe ratio-maximizing SDF.

Turning to sub-figure (b) for bivariate-sorted portfolios, the \emph{profitability} theme dominates the heatmap, particularly following a pronounced regime shift in the late 1980s. The \emph{size} theme exhibits persistent importance throughout the sample period, reflecting the strong cross-sectional dispersion in firm size within our test assets. In contrast, \emph{accruals}, \emph{profit growth}, \emph{seasonality}, and \emph{short-term reversal} show consistently low importance over the entire sample.

Our analysis reveals that while the dominant predictive role of \emph{investment}, \emph{value}, \emph{profitability}, and \emph{size} themes remains stable over time, certain signals---particularly \emph{accruals} and \emph{quality}---exhibit heightened importance during high-volatility or low-sentiment periods. This time variation suggests dynamic shifts in return predictability patterns, which our framework successfully captures through its adaptive structure.

In summary, our signal importance analysis demonstrates that the cross-predictive SDF is primarily driven by \emph{stable, economically grounded predictors}, with negligible dependence on transient or noisy effects. These findings not only underscore the robustness and economic interpretability of our framework but also open new avenues for investigating the fundamental drivers of cross-sectional return predictability.

  % \clearpage

\subsection{ Networks in the Cross Section}
\label{sec:emp_connectedness}

To uncover the economic structure embedded in the cross-predictive matrix~$\Psi$, we interpret $\Psi$ as the adjacency matrix of a directed network across $N$ assets. This representation enables us to move beyond portfolio-level effects and examine how predictive information flows through the cross-section. That is we identify assets that function as net transmitters or receivers of signals and assessing the alignment of these linkages with economic groupings such as firm size.

Following the connectedness methodology of \citet{diebold2014network}, we compute three metrics for each asset $i$---\textit{outgoing connectedness} ($\mathrm{FROM}$), \textit{incoming connectedness} ($\mathrm{TO}$) and \textit{net connectedness} ($\mathrm{NET}$)---along with a market-level  \textit{overall network intensity} ($\mathrm{TOTAL}$). 
% These measures help pinpoint dominant sources of predictability.
% and inform the imposition of economically motivated sparsity structures on $\Psi$ to improve interpretability and out-of-sample performance.
Let $\Psi_{i,j}$ denote the predictive influence of asset~$i$ on asset~$j$. We define the network metrics as follows:
\begin{align}
	\mathrm{FROM}_i &= \sum_{\substack{j=1 \\ j \neq i}}^N |\Psi_{i,j}|, \label{eqn:from} \\
	\mathrm{TO}_j &= \sum_{\substack{i=1 \\ i \neq j}}^N |\Psi_{i,j}|, \label{eqn:to} \\
	\mathrm{NET}_k &= \mathrm{TO}_k - \mathrm{FROM}_k, \label{eqn:net} \\
	\mathrm{TOTAL} &= \frac{1}{N} \sum_{\substack{i,j=1 \\ i \neq j}}^N |\Psi_{i,j}|. \label{eqn:total}
\end{align}

Here, $\mathrm{FROM}_i$ measures the total strength of predictive signals sent from asset $i$ to others, capturing how much $i$ contributes to forecasting the returns of other assets. 
$\mathrm{TO}_j$ measures the total strength of predictive signals received by asset $j$ from all other assets, reflecting how much $j$ is influenced by the rest of the network. 
$\mathrm{NET}_k$ is the difference between incoming and outgoing connectedness, indicating whether asset $k$ is a net transmitter ($< 0$) or net receiver ($> 0$) of predictive information. 
$\mathrm{TOTAL}$ aggregates the overall off-diagonal magnitude of $\Psi$ across all asset pairs, summarizing the average intensity of cross-asset predictive linkages in the network.
The use of absolute values follows \citet{diebold2014network} and ensures all measures are non-negative, thereby capturing signal strength regardless of sign.

%%%%%%% how to run the regression. %%%%%%%

We compute these metrics monthly for two asset universes—138 spread portfolios and 544 bivariate sorted portfolios—over $T = 611$ months. To investigate the firm-level characteristics driving variation in connectedness, we estimate monthly cross-sectional regressions:
\begin{equation}
	\mathrm{Connectedness}_{i,t} = \alpha_t + \boldsymbol{\beta}' \mathrm{Char}_{i,t} + \varepsilon_{i,t}, \label{eqn:cs_reg}
\end{equation}
where $\mathrm{Connectedness}_{i,t}$ is one of $\mathrm{FROM}_i$, $\mathrm{TO}_i$, or $\mathrm{NET}_i$, and $\mathrm{Char}_{i,t}$ is a vector of observable characteristics. We report time-series averages of the estimated coefficients along with Newey--West \citep{neweywest1987} $t$-statistics using a Bartlett kernel and lag length $L = 4(T/100)^{2/9} \approx 5$.

%%%%%%% overall summary of the regression results. %%%%%%%

Table~\ref{tab:psi_regression} reports the results of monthly cross-sectional regressions of three network connectedness measures—$\mathrm{FROM}$, $\mathrm{TO}$, and $\mathrm{NET}$—on firm characteristics for two groups of test assets: spread portfolios (Panel A) and bivariate sorted portfolios (Panel B). The results reveal economically intuitive patterns linking a stock’s network role to size, valuation, profitability, investment, momentum, and several trading frictions. 

%%%%%%% Panel A. FROM %%%%%%%

In Panel A for spread portfolios, the $\mathrm{FROM}$ regressions, measuring how much a stock helps predict others, we observe that smaller stocks (low ME), high book-to-market (BM), high profitability (OP), and high momentum (MOM) stocks tend to transmit stronger signals to others. These firms—small, value, profitable, and past winners—have greater forecasting influence, possibly because they aggregate market-wide information or drive co-movements. 
Additionally, stocks with low illiquidity (ILL) and low turnover (TRN) exhibit higher FROM, suggesting that liquidity increase a stock’s impact to the network. Volatility (VLT), by contrast, enters positively, implying that more volatile stocks spill predictive attention. Notably, the coefficient on size (ME) becomes insignificant, once controlling five trading frictions, which means that the size effect on $\mathrm{FROM}$ is a manifestation of trading frictions but not size itself.

%%%%%%% Panel A. TO %%%%%%%

The $\mathrm{TO}$ regressions, which capture how strongly a stock is predicted by others, show the opposite patterns on many characteristics.
Stocks with high ME, high BM, low OP, low INV, high MOM, high VLT, and high BETA receive more predictive inputs from others.
This suggests that firms that are large, volatile, illiquid, and priced as value stocks appear more susceptible to being forecasted using cross-asset information. 
Interestingly, high-MOM stocks both receive and transmit signals, indicating they may act as informational amplifiers within the network.

%%%%%%% Panel A. NET %%%%%%%

The $\mathrm{NET}$ regressions, defined as $\mathrm{TO} - \mathrm{FROM}$, consolidate these effects to identify whether a stock is a net receiver or transmitter of predictive information. Stocks that are large (high-ME), low-BM, low-OP, low-INV, and low-MOM tend to be net receivers, while small, value, profitable, non-investing, and momentum-driven stocks are net transmitters. 
These directional patterns highlight a persistent asymmetry: 
small, value, strong profitability, and conservative investment firms disseminate predictive signals, 
while larger and illiquid firms absorb them. 
% The NET regressions also reveal that turnover (TRN) consistently distinguishes transmitters from receivers, suggesting that actively traded stocks play a key role in receiving predictive information.

%%%%%%% Panel B. NET %%%%%%%

In Panel B for bivariate sorted portfolio, these patterns still exist. For ease of interpretation, we focus on the $\mathrm{NET}$ regressions. We find that small, low-BM, high-OP, low-INV, and low-MOM firms are net receivers in the network, while big, value, weak-profitable, conservative-investing, and high-momentum stocks are net transmitters.
After controlling five trading frictions in the regressions, the coefficient on size become significantly positive, while other four coefficients are unchanged.
As for trading frictions, stocks with low volume, low volatility, high turnover, and low market-beta tend to receive spillovers from others than transmitting signals to others.

%%%%%%% Panel A and B. Summary %%%%%%%

Together, two sets of test assets demonstrate significant correlations between network connectedness and asset characteristics, shedding light on that the determinants of cross-asset spillover effects. The estimated $\Psi$ matrix embeds an economically interpretable hierarchy of signal flows, shaped by firm fundamentals and market frictions. This structure supports imposing sparsity or blockwise restrictions to enhance interpretability and control overfitting—especially by limiting signal flows that contradict observed economic asymmetries. 
Nevertheless, the correlation between connectedness and asset characteristics depends on the choice of test assets. That is, different test assets reflect different patterns in asset pricing, see \cite{feng2020taming, avramov2025sparse}.
In this specific exercise, we confirm the prominent status of size as an asset characteristic in building sorted portfolios as test assets \citep{fama1993common}. 

%%%%%%% Panel A and B, link to the literature. %%%%%%%

Table~\ref{tab:psi_regression} connects to several literature.
For bivariate portfolios (Panel B), we initially corroborate \cite{lo1990contrarian}, finding big stocks lead small stocks (coefficient -0.15, row 1 on $\mathrm{NET}$)—a result robust to controlling for BM, OP, INV, and MOM (coefficient -0.19, row 2). However, controlling for trading frictions reverses the size coefficient, suggesting big stocks become net receivers, warranting further investigation of size's role in lead-lag effects.\footnote{For comparability, we replicate results for 1973-1987 (matching \cite{lo1990contrarian}'s sample end) and find consistent size coefficient signs.} 
Contrary to \cite{chordia2000trading}, we find low-turnover stocks transmit signal to high-turnover stocks after controlling for size.\footnote{While \cite{chordia2000trading} uses "Trading Volume" in their title, they actually employ daily turnover as their volume proxy.} It holds for both spread and bivariate sorts. 
The divergence from prior papers reflects discretion in test assets and sample periods. Moreover, the two papers focus exclusively on weekly return spillovers, whereas we incorporate multiple firm-level monthly signals, including past returns. Collectively, we demonstrate that cross-asset spillovers are fundamentally linked to asset characteristics.

%%%%%%% total %%%%%%%

\begin{figure}[h!]
	\caption{Total Connectedness}
	\label{fig:total_connectedness}
	{\footnotesize
		
		This figure depicts the time-series plot of total connectedness of $\Psi$ matrix over OOS period form February 1973 to December 2023. The blue dash line is for 138 spread portfolios, and the orange solid line is for 544 bivariate sorted portfolios.
		The shadow areas indicate for NBER recession periods.
	}
	\begin{center}

        \includegraphics[width=\textwidth]{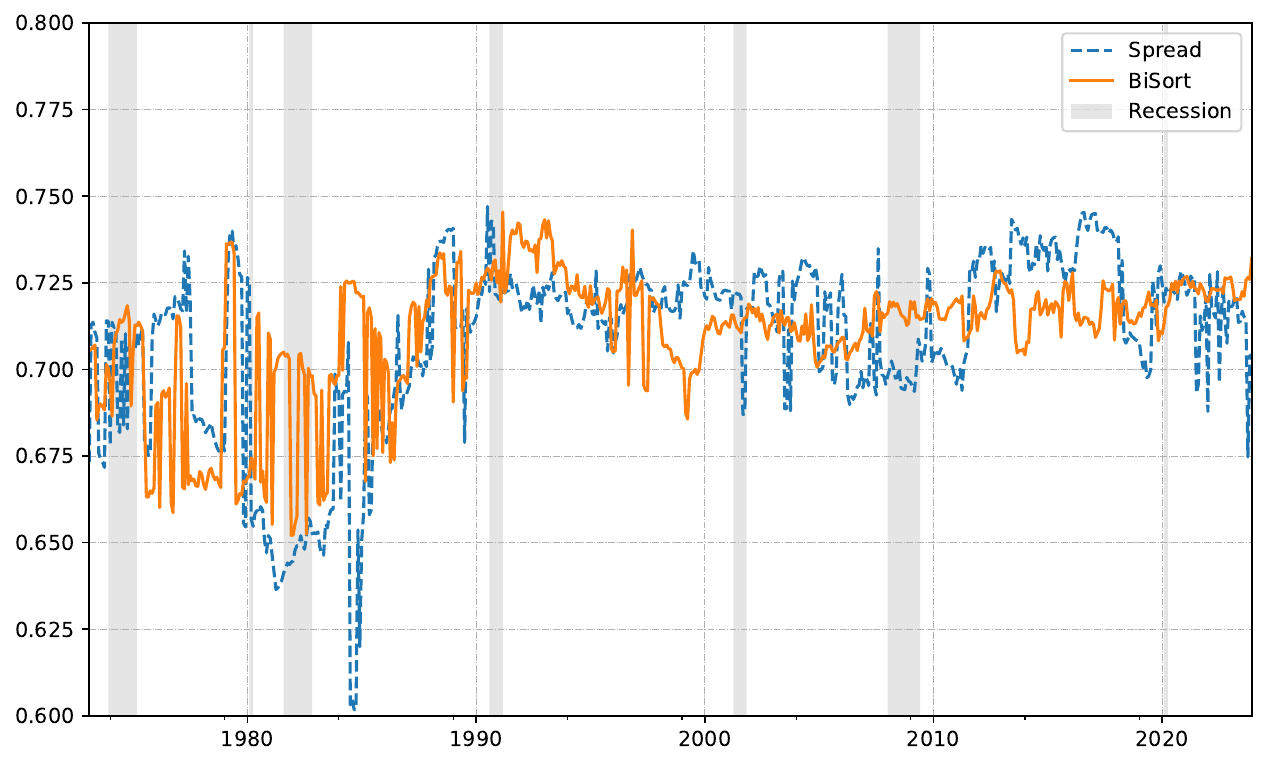}  
        
	\end{center}
	
\end{figure}

Figure~\ref{fig:total_connectedness} depicts the $\mathrm{TOTAL}$ connectedness index---the average intensity of the off-diagonal elements in $\Psi$---for both the 138 spread portfolios (dashed line) and the 544 bi-sort portfolios (solid line) over the 1973--2023 period. Four key findings emerge.  
First, the time-series of $\mathrm{TOTAL}$ connectedness on the spread portfolios varies markedly through time: it troughs in the mid-1980s and again after 2020, but peaks around the early 1990s and during the post  financial crises, 2010s. 
Second, the indices for bivariate sorted portfolios share the trough in mid-1980s and peak in early 1990s, however, slight fluctuations after 2000. 
Overall, the average level of $\mathrm{TOTAL}$ of spread portfolios is almost equal to that of bivariate sorted portfolios before 2000, but become higher after 2000.
Third, despite these episodic surges, the time series reverts to a long-run mean near 0.72, suggesting a stable baseline level of cross-asset information transmission.  
 
Taken together, Figure~\ref{fig:total_connectedness} demonstrates that cross-asset spillover effects intensify during turbulent periods but persist as a pervasive market feature. These findings underscore the importance of modeling the full $\Psi$ matrix---rather than restricting attention to its diagonal elements---for constructing Sharpe ratio-maximizing portfolios.

\clearpage

% Table generated by Excel2LaTeX from sheet 'tex'
\begin{table}[htbp]
	\caption{Understanding Connectedness}
	\label{tab:psi_regression}%
	
	{\footnotesize
		
        This table reports the time-series average and the \cite{neweywest1987} $t$-statistics of cross-sectional regressions estimates for each month that regress a connectedness measure on asset characteristics. The assets are 138 spread portfolios in Panel A, and 544 bivariate sorted portfolios in Panel B.
		For ease of interpretation, the coefficients are reported with values multiplied by 1000.
		There are three connectedness measures: $\mathrm{FROM}$, $\mathrm{TO}$, and $\mathrm{NET}$.
		The characteristics of interest are size (``market\_equity''), book-to-market equity (``be\_me''), operating profits-to-lagged book equity (``ope\_bel1''), asset growth (``at\_gr1''), price momentum t-12 to t-1 (``ret\_12\_1''), 
		Amihud illiquidity(``ami\_126d''),
		volume(``dolvol\_126d''),
		volatility(``rvol\_21d''),
		turnover(``turnover\_126d''),
		CAPM beta (``beta\_60m''),
		with abbreviations ME, BM, INV, OP, MOM, ILL, VLM, VLT, TRN, and BETA.
		There is an intercept in the regression, but the estimates are omitted in the table.
		The sample period is from February 1973 to December 2023.
	
    \vspace{-.4cm}	
    \begin{center}
    \resizebox{.85\textwidth}{!}{
    \begin{tabular}{lcccccccccc}
    \toprule
    \\
    & ME    & BM    & OP    & INV   & MOM   & ILL   & VLM   & VLT   & TRN   & BETA \\
    \\
    \hline
    \\
    \multicolumn{11}{c}{Panel A: Spread Portfolio} \\
    % \\
    \multirow{6}[0]{*}{FROM} & -1.26 &       &       &       &       &       &       &       &       &  \\
    & (-9.39) &       &       &       &       &       &       &       &       &  \\
    & -2.62 & 7.78  & 4.75  & 0.80   & 4.83  &       &       &       &       &  \\
    & (-7.62) & (16.82) & (8.39) & (3.86) & (20.28) &       &       &       &       &  \\
    & 0.37  & 8.9   & 7.04  & 0.41  & 6.47  & -7.49 & -7.04 & 4.68  & -2.49 & -0.07 \\
    & (0.16) & (27.91) & (14.08) & (1.66) & (21.42) & (-1.35) & (-1.5) & (7.36) & (-2.58) & (-0.19) \\
    \\
    \multirow{6}[0]{*}{TO} & 0.72  &       &       &       &       &       &       &       &       &  \\
    & (3.19) &       &       &       &       &       &       &       &       &  \\
    & 1.65  & 0.90   & -2.79 & -1.66 & 2.63  &       &       &       &       &  \\
    & (7.22) & (3.27) & (-7.43) & (-9.71) & (7.77) &       &       &       &       &  \\
    & 17.47 & 2.84  & -1.06 & -1.77 & 4.03  & -4.29 & -15.46 & 5.55  & -1.90  & 2.03 \\
    & (8.3) & (11.55) & (-3.01) & (-10.42) & (9.79) & (-1.11) & (-3.55) & (13.35) & (-2.14) & (5.38) \\
    \\
    \multirow{6}[0]{*}{NET} & 1.98  &       &       &       &       &       &       &       &       &  \\
    & (6.45) &       &       &       &       &       &       &       &       &  \\
    & 4.27  & -6.88 & -7.53 & -2.46 & -2.20  &       &       &       &       &  \\
    & (9.78) & (-14.07) & (-11.09) & (-11.32) & (-6.05) &       &       &       &       &  \\
    & 17.09 & -6.06 & -8.10  & -2.19 & -2.44 & 3.19  & -8.42 & 0.86  & 0.58  & 2.11 \\
    & (4.36) & (-19.12) & (-13.52) & (-8.65) & (-4.45) & (0.49) & (-1.42) & (1.06) & (0.38) & (4.22) \\
    \\
    \hline
    \\
    \multicolumn{11}{c}{Panel B: BiSort Portfolio} \\
    % \\
    \multirow{6}[0]{*}{FROM} & 0.25  &       &       &       &       &       &       &       &       &  \\
    & (12.53) &       &       &       &       &       &       &       &       &  \\
    & 0.36  & 0.38  & -0.30  & 0.59  & 0.52  &       &       &       &       &  \\
    & (13.77) & (16.23) & (-16.54) & (20.59) & (20.31) &       &       &       &       &  \\
    & 0.32  & 0.53  & -0.16 & 0.69  & 0.73  & 1.39  & 2.22  & 0.56  & -0.93 & -0.03 \\
    & (1.21) & (18.89) & (-6.22) & (19.2) & (15.16) & (2.36) & (2.98) & (7.82) & (-4.57) & (-0.41) \\
    \\
    \multirow{6}[0]{*}{TO} & 0.10   &       &       &       &       &       &       &       &       &  \\
    & (1.66) &       &       &       &       &       &       &       &       &  \\
    & 0.17  & -0.26 & 0.07  & -0.33 & -0.06 &       &       &       &       &  \\
    & (2.86) & (-4.79) & (0.95) & (-5.14) & (-0.85) &       &       &       &       &  \\
    & 3.82  & -0.05 & 0.41  & -0.45 & -0.04 & 0.77  & -3.96 & -0.23 & 2.34  & -0.50 \\
    & (8.13) & (-0.68) & (5.21) & (-6.84) & (-0.54) & (0.7) & (-3.99) & (-1.65) & (10.22) & (-3.7) \\
    \\
    \multirow{6}[0]{*}{NET} & -0.15 &       &       &       &       &       &       &       &       &  \\
    & (-2.41) &       &       &       &       &       &       &       &       &  \\
    & -0.19 & -0.64 & 0.38  & -0.92 & -0.58 &       &       &       &       &  \\
    & (-3.35) & (-11.56) & (4.54) & (-11.36) & (-10.22) &       &       &       &       &  \\
    & 3.50   & -0.59 & 0.58  & -1.14 & -0.77 & -0.63 & -6.18 & -0.79 & 3.27  & -0.47 \\
    & (6.05) & (-7.97) & (6.49) & (-13.19) & (-10.66) & (-0.63) & (-5.47) & (-5.92) & (9.36) & (-3.2) \\
    
    \\
    \bottomrule
    \end{tabular}%
    }
    \end{center}
		
	}
	
\end{table}%

\clearpage

%%%%%%% Big vs Small %%%%%%%

% \clearpage
    
To analyze directional spillover effects in the bivariate-sorted portfolios more precisely, we decompose the $\Psi$ matrix into four blocks (A, B, C, and D) according to firm size. Figure~\ref{fig:eco_bps} presents the resulting predictive information flows across these partitions.

\begin{figure}[htbp]
	\caption{Absolute Average of Four Blocks in $\Psi$: BiSort Portfolios}
	\label{fig:abcd}
	{\footnotesize
		
		This figure shows the time-series plot of the absolute average of elements in four blocks of $\Psi$. 
		The basic assets are the bivariate sorted portfolios, where four blocks A, B, C, and D represent the strength of cross-predictive relations for 
		(1) small stock signals predict small stock returns, 
		(2) small stock signals predict big stock returns, 
		(3) big stock signals predict small stock returns, 
		and
		(4) big stock signals predict big stock returns.
		The sample period is form February 1973 to December 2023.
		The shadow areas indicate for NBER recession periods.
	}
	\begin{center}
		\centering
		\includegraphics[width=\textwidth]{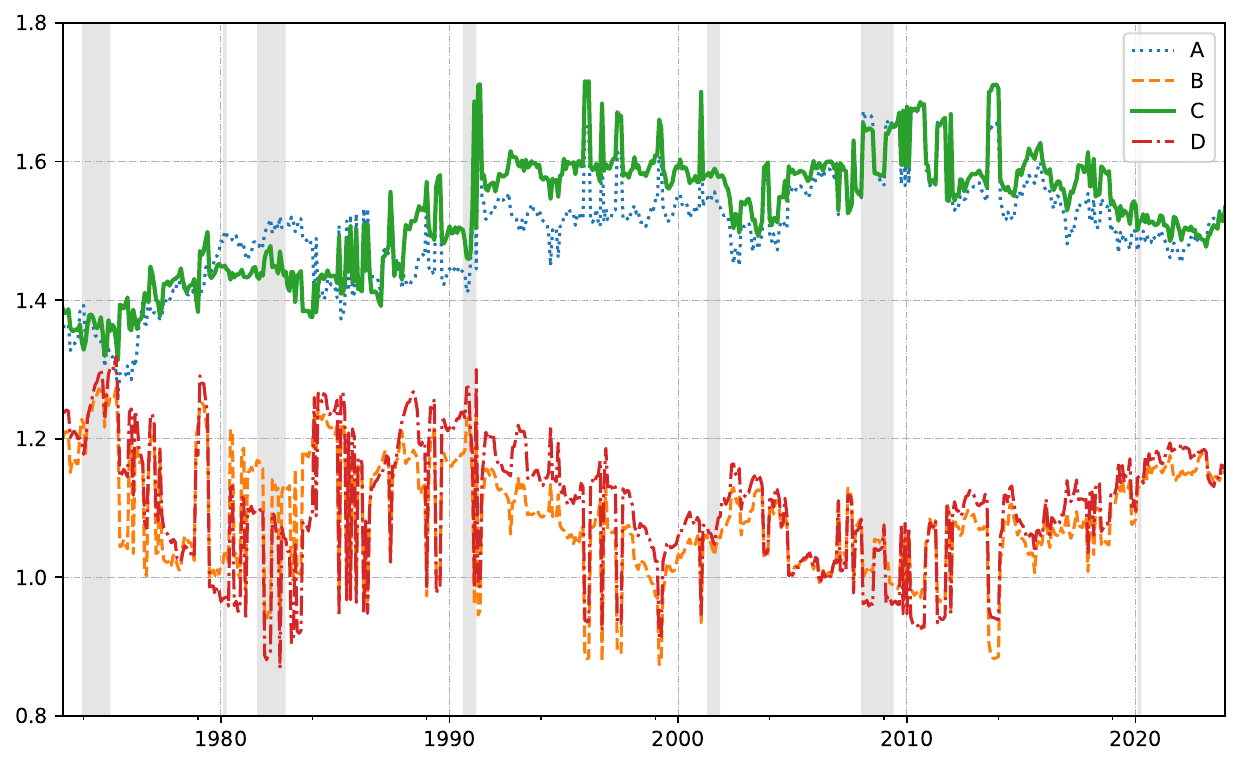}  
	\end{center}
	
\end{figure}

Figure~\ref{fig:abcd} presents the time series of absolute average values for each of the four blocks in $\Psi$. 
The results reveal consistently stronger predictive relations in Block A (Small $\rightarrow$ Small) and Block C (Big $\rightarrow$ Small) compared to Block B (Small $\rightarrow$ Big) and Block D (Big $\rightarrow$ Big), particularly during the last two decades. The time-series averages are 1.50 and 1.53 for Blocks A and C, respectively, versus 1.09 and 1.11 for Blocks B and D. Notably, the divergence between the A/C and B/D blocks has increased substantially in recent years.

These findings confirm an asymmetric predictive structure, which aligns with the $\mathrm{NET}$ regression coefficient of $-0.15$ reported in Panel B of Table~\ref{tab:psi_regression}. This result is consistent with the evidence in \citet{lo1990contrarian}, showing that large stocks tend to lead small stocks, but not vice versa. The persistent and stable nature of these patterns over time supports the economic rationale for imposing restrictions on $\Psi$, particularly by excluding small-to-large predictive links. Furthermore, the long-run regularity of these asymmetries suggests that dynamic sparsity structures---which adapt to time-varying network block strengths while maintaining economically motivated constraints---could offer significant modeling value.

In summary, the connectedness analysis reveals that the connection matrix $\Psi$ encodes economically meaningful structure. For bivariate sorted portfolios on size and other signals, big stocks act as net transmitters of predictive signals; controlling more signals, we find that low trading volume, high turnover ratio, and low-beta stocks are net transmitters. Meanwhile, value, profitable, non-investing, and high-momentum assets are more likely to be net receivers. The strength of cross-predictive relations is comparable to that of self-predictive effects. The overall network intensity fluctuates over time, but remains around a stable level. Decomposing~$\Psi$ by firm size shows that predictive flows from large to small firms dominate those in the reverse direction. 

% \clearpage

\begin{figure}[h!]
    \caption{Partition of $\Psi$ in Size.}
    \label{fig:eco_bps}

    {\footnotesize
    This figure decomposes the $\Psi$ matrix to four blocks based on firm size. They are:
    \begin{itemize}
    	\item A: Small (Stock Signals) $\rightarrow$ Small (Stock Returns),
    	\item B: Small $\rightarrow$ Big,
    	\item C: Big $\rightarrow$ Small,
    	\item D: Big $\rightarrow$ Big.
    \end{itemize}
    }
    
    \begin{center}
    
        \begin{tikzpicture}[scale=2.5]
        % Draw the first 2-by-2 grid
        \draw (0,0) grid (2,2);
        % Fill cells A, B, and D with dense dash lines
        % \fill[pattern=dots] (0,1) rectangle (1,2); % Cell A
        % \fill[pattern=dots] (1,1) rectangle (2,2); % Cell B
        % \fill[pattern=dots] (0,0) rectangle (1,1); % Cell C
        % \fill[pattern=dots] (1,0) rectangle (2,1); % Cell D
        % Label the cells (optional)
        \node at (0.5, 1.5) {\Large A};
        \node at (1.5, 1.5) {\Large B};
        \node at (0.5, 0.5) {\Large C};
        \node at (1.5, 0.5) {\Large D};
        % Add "Big Signal" to the left of node A
        \node[left]  at (0,   1.5) {Small Signal};
        \node[left]  at (0,   0.5) {Big Signal};
        \node[above] at (0.5, 2.0) {Small Return};
        \node[above] at (1.5, 2.0) {Big Return};
        \end{tikzpicture}
        
    \end{center}
    
    \end{figure}

\clearpage

\clearpage

	\clearpage	
	\section{Conclusion}
    \label{sec:conclusion}

   This paper develops a structured framework for constructing Sharpe ratio–maximizing investment strategies using multiple firm-level signals and accounting for informational linkages across assets. By jointly estimating signal relevance and a matrix capturing cross-asset predictive relationships, our approach yields closed-form portfolio weights derived from a generalized eigenvalue decomposition. In high-dimensional settings, estimation is implemented through Ridge-SDF regressions, which offer a stable and interpretable managed-portfolio representation of the decision variables. The resulting stochastic discount factor consistently delivers high out-of-sample Sharpe ratios across a range of asset universes and market conditions, outperforming both self-predictive models and expected return-maximization. Economically, the strategy is primarily driven by fundamental characteristics related to investment, valuation, and profitability. In addition, the estimated connection matrix reveals that large and low-turnover stocks tend to act as net transmitters of predictive signals, while the overall strength of cross-asset linkages remains persistently high over time.
   
  The paper opens several promising avenues for future research. First, the framework could be extended to other asset classes where cross-asset interdependencies are economically meaningful, such as corporate bonds, currencies, sovereign credit, or international equities. For instance, in the corporate bond market, issuer fundamentals or equity-side information may predict bond returns through industry linkages, shared ownership networks, or common analyst coverage. Similarly, in currency markets, major reserve currencies may act as informational hubs whose movements help forecast subsequent shifts in peripheral currencies. Second, incorporating economic structure into the modeling of cross-asset relationships could enhance both interpretability and predictive performance. As the number of assets and signals expands, estimating all possible interactions becomes increasingly challenging. Imposing economically motivated constraints—such as directional spillovers based on firm size or sectoral hierarchies—could provide a more structured and scalable approach.

	%%%%%%%%%%%%%%%%%%%%%%%%%%
	% Reference %
	%%%%%%%%%%%%%%%%%%%%%%%%%%

    \clearpage
    
	\bibliography{cite.bib}
	
	%%%%%%%%%%%%%%%%%%%%%%%%%%
	% More Tables and Figures %
	%%%%%%%%%%%%%%%%%%%%%%%%%%

	%%%%%%%%%%%%%%%%%%%%%%%%%
	\clearpage
	\setstretch{1.6}
	\appendix
	\appendixpage
	%%%%%%%%%%%%%%%%%%%%%%%%%%
	% Appendix %
	%%%%%%%%%%%%%%%%%%%%%%%%%%

	\renewcommand{\thesection}{\Alph{section}}
	\renewcommand{\thesubsection}{\Alph{section}.\arabic{subsection}}
	\renewcommand{\thesubsubsection}{\Alph{section}.\arabic{subsection}.\arabic{subsubsection}}
	\makeatother
	
	\renewcommand\theequation{\Alph{section}.\arabic{equation}}
	\renewcommand\thefigure{\Alph{section}.\arabic{figure}}
	\renewcommand\thetable{\Alph{section}.\arabic{table}}
    
    \makeatletter
    \@addtoreset{equation}{section}
    \@addtoreset{figure}{section}
    \@addtoreset{table}{section}
    \makeatother

	\section{Proof of Proposition \ref{prp:linear_metrics} } 	\label{app:linear_metrics}
	
	\paragraph{Expected Return}
	We first express \( \pi_s \), the realized return on the trading strategy, as a function of the model parameters.
	Recognize that \(\pi_s = \Lambda' S'_t \Psi r_s = \sum_{i=1}^{N} \Psi_i' r_s S_{it}' \Lambda = \text{tr}\left[\Lambda \sum_{i=1}^{N} \Psi_i' \Pi_{si}\right] = \Lambda' \Pi_s' \Phi\), where \(\Psi_i'\) is a \(1 \times N\) vector which is the \(i\)-th row of \(\Psi\),  \(S_{it}'\) is a \(1 \times M\) vector, which is  the \(i\)-th row of \(S_t\), \(\text{tr}\) stands for the trace operator, and \(\Pi_s\) is an \(N^2 \times M\) matrix that vertically stacks the \(N \times M\) matrices \(\Pi_{si} = r_s S_{it}'\) for \(i = 1, 2, \cdots, N\).

	Then, on the basis of realized return, the expected value is given by
	\begin{equation}
		E(\pi_s) = \Lambda' {\Pi}' \Phi = \Phi' {\Pi} \Lambda.
	\end{equation}
	
	\paragraph{Variance}	
 Let $\Sigma_{\Phi}$ be the covariance matrix of \(\text{vec}(\Pi_s')\). We express \(\pi_s\) in terms of \(\text{vec}(\Pi_s')\):
	\begin{equation}
		\pi_s = \Lambda' \Pi_s' \Phi = \Lambda' \text{vec}(\Pi_s' \Phi).
	\end{equation}
	
	Using the property of vectorization $
		\text{vec}(ABC) = (C' \otimes A) \text{vec}(B)$,
	we get:
	\begin{equation}
		\text{vec}(\Pi_s' \Phi) = (\Phi' \otimes I_M) \text{vec}(\Pi_s').
	\end{equation}
	
	Therefore:
	\begin{equation}
		\pi_s = \Lambda' (\Phi' \otimes I_M) \text{vec}(\Pi_s').
	\end{equation}
	
	The variance of \(\pi_s\) is:
	\begin{eqnarray}\nonumber
		\text{Var}(\pi_s) &=& \Lambda' (\Phi' \otimes I_M) \Sigma_{\Phi} (\Phi \otimes I_M) \Lambda, \\
		&=& \Lambda' B_{\Phi} \Lambda,
	\end{eqnarray}
	where $B_{\Phi} = (\Phi' \otimes I_M) \Sigma_{\Phi} (\Phi \otimes I_M).$
	
	We consider an alternative expression of \(\text{Var}(\pi_s)\).  Let $\Sigma_{\Lambda}$ be the covariance matrix of \(\text{vec}(\Pi_s)\). We express \(\pi_s\) in terms of \(\text{vec}(\Pi_s)\):
	\begin{equation}
		\pi_s = \Phi' \Pi_s \Lambda = \Phi' \text{vec}(\Pi_s \Lambda).
	\end{equation}
	
	Again using the property of vectorization, we get:
	\begin{equation}
		\text{vec}(\Pi_s \Lambda) = (\Lambda' \otimes I_{N^2}) \text{vec}(\Pi_s).
	\end{equation}
	
	Therefore:
	\begin{equation}
		\pi_s = \Phi' (\Lambda' \otimes I_{N^2}) \text{vec}(\Pi_s).
	\end{equation}		
	
	The variance of \(\pi_s\) is:
	\begin{eqnarray}\nonumber
		\text{Var}(\pi_s) &=& \Phi' (\Lambda' \otimes I_{N^2}) \Sigma_{\Lambda} (\Lambda \otimes I_{N^2}) \Phi, \\
		&=& \Phi' B_{\Lambda} \Phi,
	\end{eqnarray}
	where $B_{\Lambda} = (\Lambda' \otimes I_{N^2}) \Sigma_{\Lambda} (\Lambda \otimes I_{N^2}).$
	
	\paragraph{Sharpe Ratio}
	With the expected return and variance, we express the Sharpe ratio square as:
	\begin{equation}
		\label{eqn:sr2_Lambda}
		S R^2 = 
		\frac{\Lambda^{\prime} A_{\Phi} \Lambda}{\Lambda^{\prime} B_{\Phi} \Lambda},
	\end{equation}
	where $A_{\Phi} = \Pi^\prime \Phi \Phi^\prime \Pi, B_{\Phi}=\left(\Phi^{\prime} \otimes I_M\right) \Sigma_{\Phi} \left(\Phi \otimes I_M\right)$, and $\Sigma_\Phi$ is the covariance matrix of $\text{vec}(\Pi_s^\prime)$.		
	Alternatively, we express the Sharpe ratio squared as:
	\begin{equation}
		\label{eqn:sr2_Phi}
		S R^2 = 
		\frac{\Phi^{\prime} A_{\Lambda} \Phi}{\Phi^{\prime} B_{\Lambda} \Phi},
	\end{equation}
	where  $A_{\Lambda} = \Pi \Lambda \Lambda^\prime \Pi^\prime, B_{\Lambda}=\left(\Lambda^{\prime} \otimes I_{N^2}\right) \Sigma_{\Lambda} \left(\Lambda \otimes I_{N^2}\right)$, and $\Sigma_\Lambda$ is the covariance matrix of  $\text{vec}(\Pi_s)$. These alternative expressions of \( SR^2 \) assist in finding the solution to maximize the Sharpe ratio.
	
	% \clearpage

	\section{Relating $\Phi$ to $B$ When $M = 1$}
	\label{sec:dgp}
	
	\paragraph{Setup.}
    Consider the return-generating process:
	\begin{equation}
		r_s = B S_t + \varepsilon_s,
	\end{equation}
	where:
	\begin{itemize}
		\item $S_t \in \mathbb{R}^{N \times 1}$ is the signal vector,
		\item $B \in \mathbb{R}^{N \times N}$ is the slope matrix,
		\item $\varepsilon_s \sim (0, \Sigma_\varepsilon)$ is a zero-mean innovation,
		\item $\Sigma_S = \mathbb{E}[S_t S_t']$ is the signal covariance matrix.
	\end{itemize}
	
	\paragraph{Managed-Portfolio Return.}
	
	The managed-portfolio return vector is defined as:
	\begin{equation}
		\Pi_s = (I_N \otimes r_s) S_t.
	\end{equation}
	Taking expectations:
	\begin{equation}
		\Pi = \mathbb{E}[\Pi_s] = \text{vec}(\mathbb{E}[r_s S_t']) = \text{vec}(B \Sigma_S).
	\end{equation}
	In the case of a single signal, both $\Pi$ and $\Pi_s$ are vectors of dimension $N^2$, and $\Sigma_\Lambda$ denotes the covariance matrix of $\Pi_s$.

	\paragraph{Sharpe Ratio Maximization.}
	
	We maximize the Sharpe ratio subject to $\|\Phi\| = 1$:
	\begin{equation}
		\max_{\Phi: \|\Phi\| = 1} \frac{\Phi' \Pi}{\sqrt{\Phi' \Sigma_\Lambda \Phi}}.
	\end{equation}
	The optimal solution is:
\begin{equation}
		\Phi = \frac{\Sigma_\Lambda^{-1} \text{vec}(B \Sigma_S)}{\| \Sigma_\Lambda^{-1} \text{vec}(B \Sigma_S) \|}.
	\end{equation}
	Thus, we obtain:
	\begin{equation}
		B = \text{unvec}(\Sigma_\Lambda \Phi) \cdot \Sigma_S^{-1},
	\end{equation}
where \( \operatorname{unvec} \) denotes the reshaping of an $N^2$-vector into an \( N \times N \) matrix.

	\paragraph{Expected Return Maximization.}
	
	We now maximize expected return subject to $\|\Phi\| = 1$:
	\begin{equation}
		\max_{\Phi: \|\Phi\| = 1} \Phi' \text{vec}(B \Sigma_S)
	\end{equation}
	The solution is:
	\begin{equation}
		\Phi = \frac{\text{vec}(B \Sigma_S)}{\| \text{vec}(B \Sigma_S) \|}.
	\end{equation}
	Inverting gives:
	\begin{equation}
		B = \text{unvec}(\Phi) \cdot \Sigma_S^{-1}.
	\end{equation}

	\paragraph{Intuition: Why Are the $B$ Matrices Different?}
	
	The difference stems from the objective:
	\begin{itemize}
		\item \textbf{Maximizing Expected Return:} aligns $\Phi$ with the direction of highest expected payoff, ignoring variance.
		\item \textbf{Maximizing Sharpe Ratio:} adjusts for risk by incorporating $\Sigma_\Lambda$, penalizing high-volatility directions.
	\end{itemize}
	
	% \clearpage
	
	\section{Proof of Expected Return Reduction due to Zero-Cost Constraint} \label{app:zero_cost}
	
	Consider the matrix \(\Pi\) formed by vertically stacking \(N\) matrices \(\Pi_i\), each of dimension \(N \times M\), and let \(\tilde{\Pi}\) be the matrix obtained after pre-multiplying each \(\Pi_i\) by the matrix \(\Theta\), where \(\Theta = I_N - \frac{1}{N} \iota_N \iota_N'\). Here, \(\Theta\) is a projection matrix that projects vectors onto the space orthogonal to the vector \(\iota_N\) of ones.
	
	\textbf{Properties of \(\Theta\):}
	\begin{itemize}
		\item \(\Theta\) is symmetric and idempotent, i.e., \(\Theta^2 = \Theta\) and \(\Theta' = \Theta\), confirming that it is a projection matrix.
		\item The eigenvalues of \(\Theta\) are 0 along the direction of \(\iota_N\) and 1 along all directions orthogonal to \(\iota_N\).
	\end{itemize}
	
	\textbf{Impact on Singular Values:}
	\begin{enumerate}
		\item The matrix \(\Theta\) modifies \(\Pi_i\) by removing its component in the direction of \(\iota_N\). This operation reduces the variance in \(\Pi_i\) that is aligned with \(\iota_N\).
		\item Given the singular value decomposition of \(\Pi = U \Sigma V'\), the transformation \(\tilde{\Pi} = (\Theta \Pi_i)\) can be viewed through the lens of modified singular vectors. Since \(\Theta\) acts as an identity on the space orthogonal to \(\iota_N\) and zeroes out components along \(\iota_N\), it does not increase the magnitude of any singular vector components.
		\item The singular values \(\lambda_i(\tilde{\Pi})\) of the transformed matrix \(\tilde{\Pi}\) correspond to the norms of the vectors \(\Theta U_i\), where \(U_i\) are the left singular vectors of \(\Pi\). Since \(\Theta\) is a projection (and thus a norm-reducing operation except where it acts as the identity), we have:
		\begin{equation}
		\|\Theta U_i\| \leq \|U_i\| .
		\end{equation}
		\item Therefore, the singular values of \(\tilde{\Pi}\) must satisfy:
	\begin{equation}
		\lambda_i(\tilde{\Pi}) \leq \lambda_i(\Pi).
	\end{equation}
		for each \(i\), because the projection does not increase vector norms and reduces them for vectors with non-zero components in the direction of \(\iota_N\).
	\end{enumerate}
	
	To be more precise, the highest singular value of the transformed matrix does not change due to the preservation of the highest singular value by $\Theta$. However, the transformation induced by $\Theta$ results in a reduction of singular values in the transformed matrix $\tilde{\Pi}$ in the other singular values, leading to a decrease in variance explained by certain components. Specifically, at least one singular value of $\tilde{\Pi}$ is strongly diminished compared to the corresponding singular value of the original matrix $\Pi$. This reduction underscores the effectiveness of the transformation in diminishing the influence of certain components in $\Pi$ and highlights its role in variance reduction. Hence, both expected return and risk of the trading strategy are lower in the presence of the zero-cost restriction.

    \section{Proof and Derivations for Propositions \ref{prp:max_sharpe_ratio} and \ref{prp:max_sharpe_ratio_ridge}}
	\label{apx:max_sharpe_ratio}
	
	This section focuses on maximizing the squared Sharpe ratio of a linear strategy. The results extend naturally to the Sharpe ratio maximization of a nonlinear strategy with an augmented signal space, for which we leave for future research.
	
	Maximizing the squared Sharpe ratio constitutes a \textit{generalized Rayleigh quotient problem}, which can be solved via an \textit{eigenvalue problem}. However, in empirical settings, the solution to this \textit{eigenvalue problem} often becomes ill-conditioned in high-dimensional settings.
	
	To address this issue, we employ Ridge-SDF regressions to estimate the decision variables, providing an intuitive managed-portfolio interpretation. Finally, we present an iterative algorithm to estimate $\Lambda$ and $\Phi$ until convergence. The details are as follows.
	
	\paragraph{Define the squared Sharpe ratio as a function of $\Lambda$.}
	According to Proposition \ref{prp:linear_metrics}, the squared Sharpe ratio takes the form:
	\begin{equation}
		S R^2 = \frac{\Lambda^{\prime} A_{\Phi} \Lambda}{\Lambda^{\prime} B_{\Phi} \Lambda},
	\end{equation}
	where $A_{\Phi} = \Pi^\prime \Phi \Phi^\prime \Pi$, $B_{\Phi} = (\Phi^{\prime} \otimes I_M) \Sigma_{\Phi} (\Phi \otimes I_M)$, and $\Sigma_\Phi$ is the covariance matrix of $\text{vec}(\Pi_s^\prime)$.
	
	\paragraph{Maximizing the squared Sharpe ratio with respect to $\Lambda$.}
	From \eqref{eqn:sr2_Lambda}, the optimization problem is formulated as:
	\begin{equation}
		\max_{\Lambda} \frac{\Lambda^\prime A_{\Phi} \Lambda}{\Lambda^\prime B_{\Phi} \Lambda}.
	\end{equation}
	This is equivalent to the constrained optimization problem:
	\begin{equation}
		\max_\Lambda \Lambda^\prime A_{\Phi} \Lambda 
		\quad \text{s.t.} \quad \Lambda^\prime B_{\Phi} \Lambda = \kappa.
	\end{equation}
	Given the norm constraint on $\Lambda$, we set $\kappa = 1$ without loss of generality.
	
	Applying the method of Lagrange multipliers, we define the Lagrangian function:
	\begin{equation}
		L(\Lambda, \lambda) = \Lambda^\prime A_{\Phi} \Lambda - \lambda (\Lambda^\prime B_{\Phi} \Lambda - 1).
	\end{equation}
	Taking derivatives with respect to $\Lambda$ yields the \textit{generalized eigenvalue problem}:
	\begin{equation}
		A_{\Phi} \Lambda = \lambda B_{\Phi} \Lambda.
	\end{equation}
	Multiplying both sides by $B_{\Phi}^{-1}$ gives:
	\begin{equation}
		B_{\Phi}^{-1} A_{\Phi} \Lambda = \lambda \Lambda.
	\end{equation}
	Defining $C_{\Phi} = B_{\Phi}^{-1} A_{\Phi}$, we obtain the standard \textit{eigenvalue problem}:
	\begin{equation}
		\label{eqn:C_eigen}
		C_{\Phi} \Lambda = \lambda \Lambda.
	\end{equation}
	Solving \eqref{eqn:C_eigen} provides the eigenvector corresponding to the largest eigenvalue, $\Lambda_{\text{max}}$. Normalizing for the norm constraint, we set:
	\begin{equation}
		\Lambda = \frac{\Lambda_{\text{max}}}{||\Lambda_{\text{max}}||}.
	\end{equation}
	Since the solution for $\Lambda$ depends on $\Phi$, we define the function:
	\begin{equation}
		\label{eqn:solve_Lambda}
		\Lambda = \argmax_{\Lambda} \frac{\Lambda^\prime A_{\Phi} \Lambda}{\Lambda^\prime B_{\Phi} \Lambda} = \mathbf{\Lambda}(\Phi).
	\end{equation}
	
	\paragraph{Estimating high-dimensional $\Lambda$ using ridge regression.}
	In high-dimensional settings where $M$ is large relative to the number of observations $T$, the solution in \eqref{eqn:solve_Lambda} often fails in out-of-sample tests.
	
	To address this, consider a set of managed-portfolios $\chi_{\Phi}$ of dimension $T \times M$:
    \begin{equation}
        \label{eqn:app_managed_portfolio_for_Lambda}
        \chi_\Phi = 
    \begin{bmatrix}
    	(\chi_\Phi)_{2}' \\
    	(\chi_\Phi)_{3}' \\
    	\vdots \\
    	(\chi_\Phi)_{T+1}' \\
    \end{bmatrix},
    \end{equation}
    where 
    \begin{equation}
    (\chi_{\Phi})_{s} = \Pi_s' \Phi. 
    \end{equation}
    
	The optimization in \eqref{eqn:solve_Lambda} is an asset allocation problem in which the goal is to determine the investment weights $\Lambda$ for the managed-portfolios $\chi_{\Phi}$ to maximize the squared Sharpe ratio. This is equivalent to estimating $\Lambda$ as the mean-variance efficient portfolio weights.
	
	Following \cite{britten1999sampling}, we estimate $\Lambda$ using the regression:
	\begin{equation}
		\label{eqn:hat_Lambda_ols}		
		\mathbf{1} = \chi_\Phi \Lambda  + \mathbf{u},
	\end{equation}
	where $\mathbf{1}$ is a $T$-vector of ones.
	
	To improve out-of-sample performance, we adopt ridge regression, as in \cite{kelly2023FML, shen2024can}:
	\begin{equation}
		\label{eqn:hat_Lambda_ridge_app}		
		\hat \Lambda  = (\chi_\Phi' \chi_\Phi + \lambda I_M)^{-1} \chi_\Phi' \mathbf{1},
	\end{equation}
	where $\lambda$ is a shrinkage parameter. The solution in \eqref{eqn:hat_Lambda_ridge_app} coincides with \eqref{eqn:solve_Lambda} when $\lambda = 0$. While \eqref{eqn:hat_Lambda_ridge_app} may underperform in-sample, it improves robustness for out-of-sample applications.
	
	\paragraph{Define the squared Sharpe ratio as a function of $\Phi$.}	
	Alternatively, we express the squared Sharpe ratio as:
	\begin{equation}
		S R^2 = 
		\frac{\Phi^{\prime} A_{\Lambda} \Phi}{\Phi^{\prime} B_{\Lambda} \Phi},
	\end{equation}
	where $A_{\Lambda} = \Pi \Lambda \Lambda^\prime \Pi^\prime$, $B_{\Lambda} = (\Lambda^{\prime} \otimes I_{N^2}) \Sigma_{\Lambda} (\Lambda \otimes I_{N^2})$, and $\Sigma_\Lambda$ is the covariance matrix of $\text{vec}(\Pi_s)$.
	
	\paragraph{Maximizing the squared Sharpe ratio with respect to $\Phi$.}
	Referring to Eq. \eqref{eqn:sr2_Phi}, we formulate the optimization problem as:
	\begin{equation}
		\label{eqn:solve_Phi}
		\Phi = \argmax_{\Phi} \frac{\Phi^\prime A_{\Lambda} \Phi}{\Phi^\prime B_{\Lambda} \Phi} = \mathbf{\Phi}(\Lambda).
	\end{equation}
	Solving \eqref{eqn:solve_Phi} follows the same procedure as \eqref{eqn:solve_Lambda}.
	
	\paragraph{Estimating high-dimensional $\Phi$ using ridge regression.}
	Analogous to \eqref{eqn:hat_Lambda_ridge_app}, we define managed-portfolios $\chi_{\Lambda}$ of dimension $T \times N^2$:
    \begin{equation}
        \label{eqn:app_managed_portfolio_for_Phi}
        \chi_\Lambda = 
    \begin{bmatrix}
    	(\chi_\Lambda)_{2}' \\
    	(\chi_\Lambda)_{3}' \\
    	\vdots \\
    	(\chi_\Lambda)_{T+1}' \\
    \end{bmatrix},
    \end{equation}
    where 
   \begin{equation}
    (\chi_\Lambda)_{s} = \Pi_s \Lambda.
    \end{equation}

    Applying ridge regression, we estimate $\Phi$ as:
	\begin{equation}
		\label{eqn:hat_Phi_ridge_app}		
		\hat \Phi  = (\chi_\Lambda' \chi_\Lambda + \lambda I_{N^2})^{-1} \chi_\Lambda' \mathbf{1}.
	\end{equation}

	\paragraph{Algorithm and Iteration.}
	
	To solve the whole problem, we do iterations until convergence. In each iteration, we have four steps:
	\begin{enumerate}
		
		\item Given $\Pi, \Sigma_\Lambda, \Lambda$, update the values of $A_{\Lambda}, B_{\Lambda}, C_{\Lambda}$,
		
		\item Solve Eq.\eqref{eqn:hat_Phi_ridge_app} to get the updated $\Phi$,
		
		\item Given $\Pi, \Sigma_\Phi, \Phi$, update the values of $A_{\Phi}, B_{\Phi}, C_{\Phi}$,
		
		\item Solve Eq.\eqref{eqn:hat_Lambda_ridge_app} to get the updated $\Lambda$.
		
	\end{enumerate}
	
	A full description of the algorithm is in Algorithm \ref{alg:maxSR}.

	% \clearpage
    \bigskip
	%%%%%%%%%%%%%%%%%%%%%%%
	\begin{algorithm}[htbp]
		\caption{\bf Maximize Sharpe Ratio} \label{alg:maxSR}
		% \bigskip
		\small
		\begin{algorithmic}[1]
			\Procedure{MaxSR}{{ $ \Lambda, \Phi$ }} \\
			\textbf{Input} Asset returns $r_{s}$ and signals $S_{t}$. \\
			\textbf{outcome} Investment decision variables $ \Lambda, \Phi$.
			\State Calculate $\Pi, \Sigma_{\Phi}, \Sigma_{\Lambda}$. \Comment{These variables are Constant.}
			\State Initialize index of iteration $k=0$. \Comment{We use $k$ in notation $\Lambda^{\{k\}}, \Phi^{\{k\}}$.}
			\State Initialize $\Lambda^{\{0\}}$. \Comment{E.g., the solution in Max Expected Return strategy.}
			\While{Termination Conditions \textit{not} Activated} 
			\State Update $A_{\Lambda}, B_{\Lambda}, C_{\Lambda}$ with $\Lambda^{\{k\}}$.
			\State Update $\Phi^{\{k+1\}}$ by solving Eq.\eqref{eqn:hat_Phi_ridge}.			
			$$
			\Phi^{\{k+1\}} = \mathbf{\Phi}(\Lambda^{\{k\}}).
			$$
			\State Update $A_{\Phi}, B_{\Phi}, C_{\Phi}$ with $\Phi^{\{k+1\}}$,
			\State Update $\Lambda^{\{k+1\}}$ by solving Eq.\eqref{eqn:hat_Lambda_ridge}.
			$$
			\Lambda^{\{k+1\}} = \mathbf{\Lambda}(\Phi^{\{k+1\}}).
			$$
			\State $k = k+1$.
			\EndWhile
			\State \textbf{return} $\Lambda^{\{k\}}, \Phi^{\{k\}}$ 
			\EndProcedure
		\end{algorithmic}
	\end{algorithm}

    \clearpage

    \section{Cross-Validation for $\lambda$ the Ridge Shrinkage Parameter}
    
    \label{sec:cross-validation}
    
\begin{figure}[h!]
    \caption{Selected Parameter by Cross-Validation}
    \label{fig:cv}

    {\footnotesize
    
    This table reports the selection results of $\lambda$ in \eqref{eqn:hat_Lambda_ridge} and \eqref{eqn:hat_Phi_ridge} via the five-fold cross-validation. The parameter grid is $[10^{4},10^{3},10^{2},10^{1},10^{0},10^{-1},10^{-2},10^{-3},10^{-4},10^{-5},10^{-6}]$. Each point in the figure represents the selected $\lambda$ for a rolling window estimation.
    
	\begin{center}

		\begin{subfigure}[b]{0.55\textwidth}
			\centering
			\includegraphics[width=\textwidth]{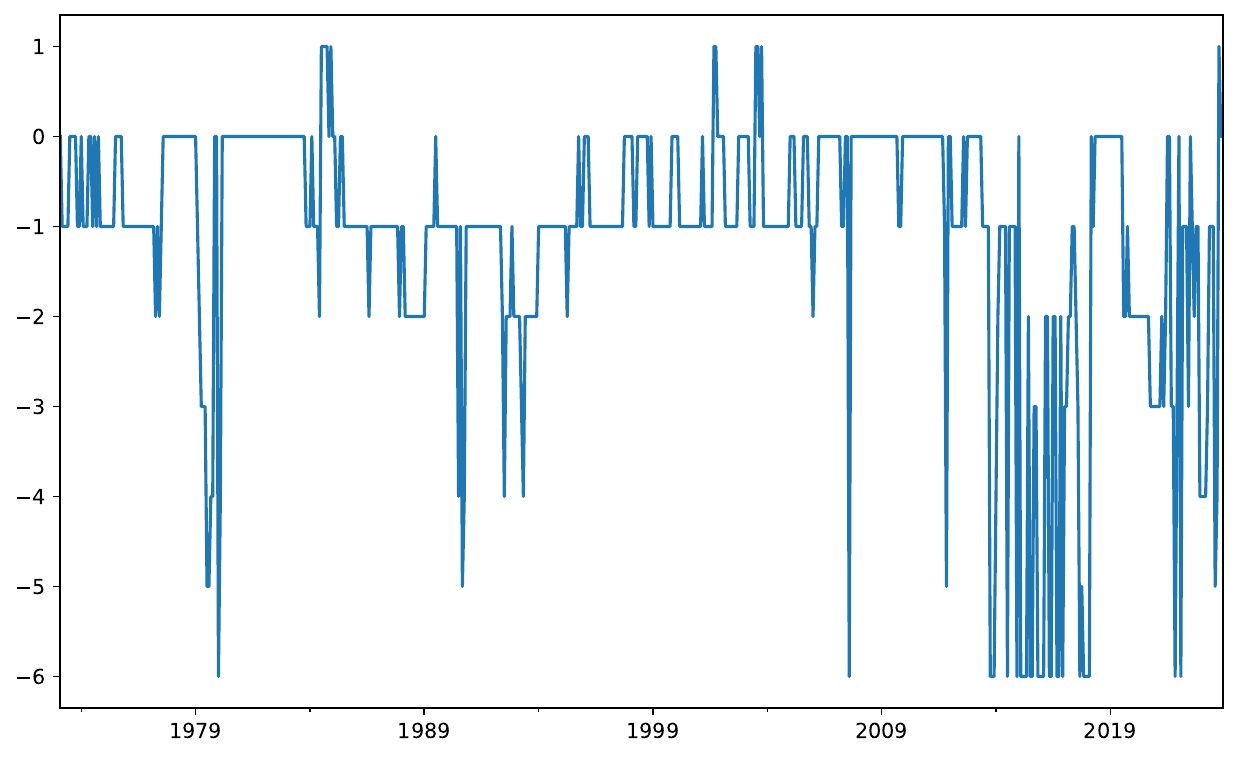}
			\caption*{(A) Spread Portfolio}
		\end{subfigure}
		
		\begin{subfigure}[b]{0.55\textwidth}
			\centering
			\includegraphics[width=\textwidth]{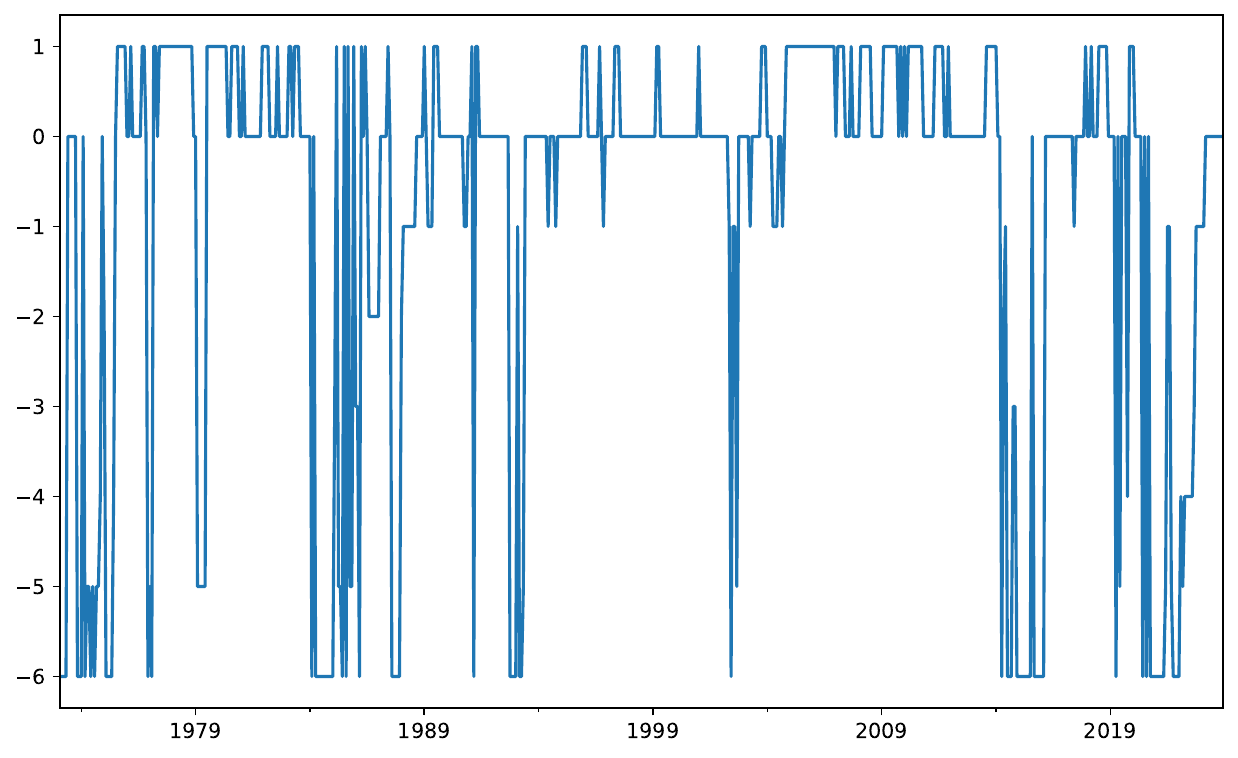}
			\caption*{(B) BiSort Portfolio}
		\end{subfigure}
    
	\end{center}

    }

    \end{figure}

    We employ the five-fold cross-validation to select the $\lambda$ parameter in ridge regressions \eqref{eqn:hat_Lambda_ridge} and \eqref{eqn:hat_Phi_ridge}, and then apply to out-of-sample investment. 
    The parameter grid is $10^{x}$, where $x\in[4,3,2,\cdots,-5,-6]$.
    Figure \ref{fig:cv} shows the parameters selected by cross-validation in each rolling window estimation.
    We find the selected parameters are time-varying, wandering in the parameter grid.

    \clearpage

    \section{Signal-Level Importance}
    \label{sec:signal_importance_app}

\begin{figure}[h!]
	\caption{Signal Importance}
	\label{fig:signal_importance_app}
	{\footnotesize

        This figure complements the theme-level signal importance in Figure \ref{fig:signal_importance} by providing the 138 signal-level importance in full detail.
        These signals of the same theme are grouped in the vertical axis, where the 13 themes follow \cite{jensen2023there}.   
        For interpretation, we focus on the absolute value of elements in $\Lambda$.
        Sub-figures (a) and (b) report for spread portfolios and bivariate sorted portfolios, respectively.

	\begin{center}

		\begin{subfigure}[b]{0.75\textwidth}
			\centering
			\includegraphics[width=\textwidth]{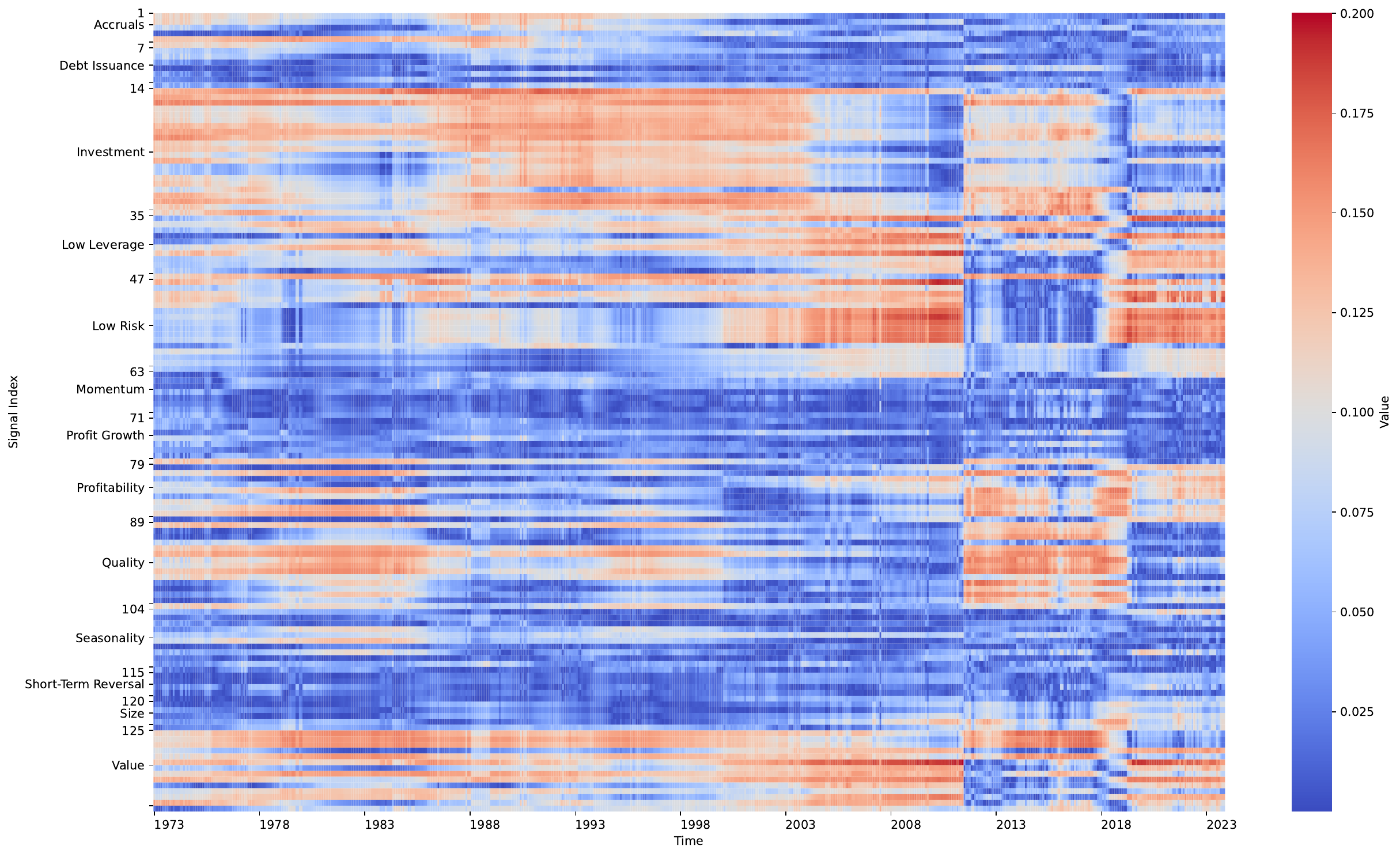}
			\caption*{(a) Spread Portfolio}
		\end{subfigure}
		
		\begin{subfigure}[b]{0.75\textwidth}
			\centering
			\includegraphics[width=\textwidth]{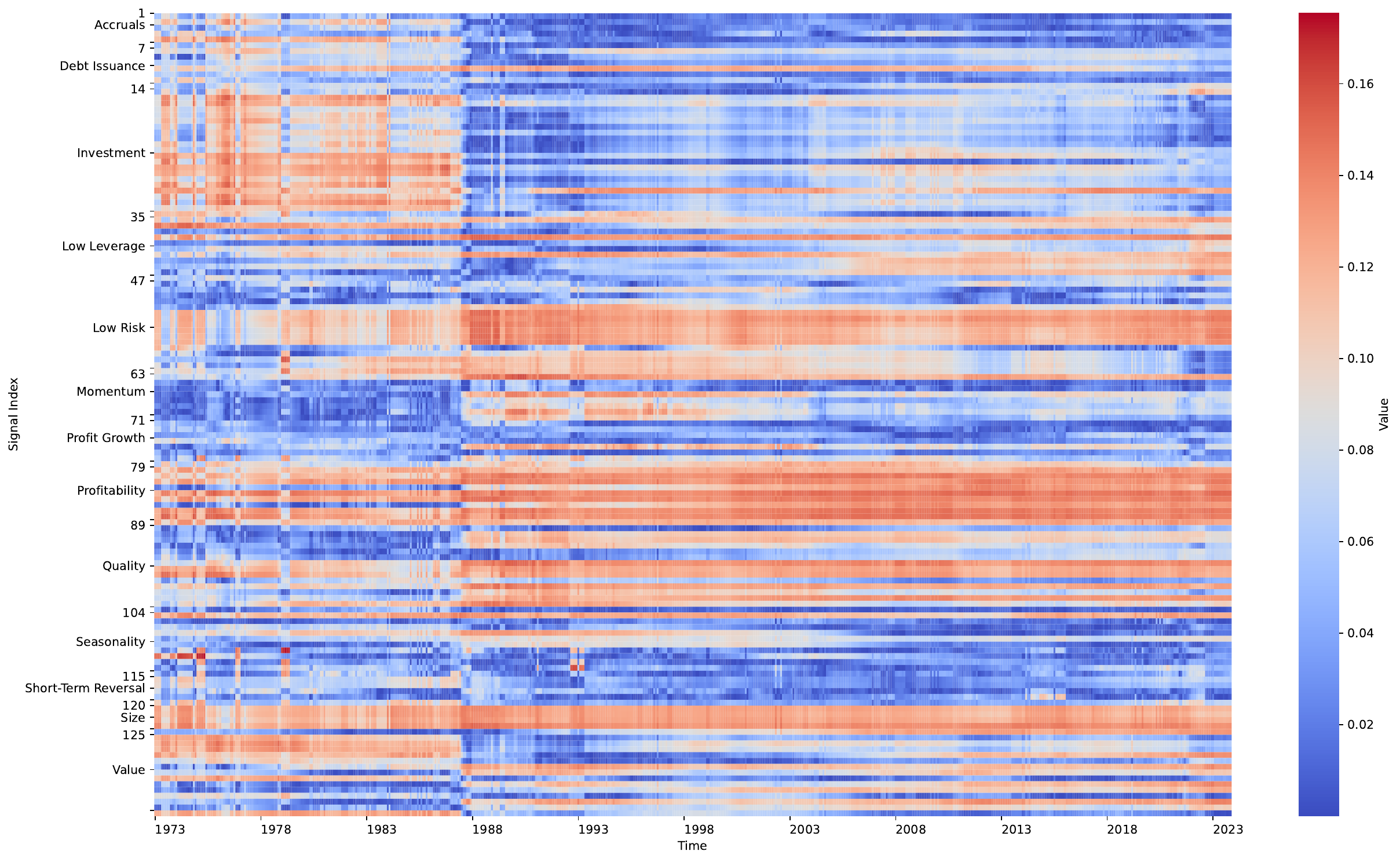}
			\caption*{(b) BiSort Portfolio}
		\end{subfigure}
		
	\end{center}

    }
	
\end{figure}

\end{document}